\documentclass[10pt]{article} 
\usepackage{graphicx}
\usepackage{hyperref}
\usepackage{amssymb}
\usepackage[italian]{babel}
\usepackage[latin1]{inputenc}
\usepackage{authblk}

\title{Neutrini in profondit\`{a}: Vita, morte e miracoli dei neutrini rivelati sotto terra, sotto i ghiacci o in fondo al mare }

\begin{document} 

\author[1]{Maurizio Spurio}
\affil[1]{{Dipartimento di Fisica e Astronomia Universit\`{a} di Bologna and // INFN Sezione di Bologna}}


\maketitle 

\vskip 1.0cm
\begin{abstract}

\noindent Il neutrino è la particella più elusiva che conosciamo e quando esso fu ipotizzato si dubitava che potesse essere mai rivelato. Oggi conosciamo e sappiamo rivelare neutrini di origine naturale (principalmente prodotti da oggetti astrofisici e dalle interazioni dei raggi cosmici) e artificiale (prodotti dai reattori nucleari e come particelle secondarie negli acceleratori). Il focus di questo articolo è sui neutini di origine naturale, per la cui rivelazione sono necessari enormi apparati sperimentali posti in laboratori sotterranei, sott'acqua o sotto il ghiaccio del Polo Sud. 
Questi studi hanno permesso enormi progressi nella conoscenza delle proprietà dei neutrini con la scoperta del meccanismo delle oscillazioni. E, nel contempo, hanno aperto nuove frontiere per lo studio dell'astrofisica dei processi che producono energia all'interno del Sole; sui meccanismi che portano al collasso gravitazionale stellare; sugli oggetti astrofisici che producono raggi cosmici sino ad energie estreme. 

\vskip 0.6cm
{\centerline{\textbf{Abstract}}}
\vskip 0.2cm

\noindent The neutrino is the most elusive particle that we know and for many years physicists doubted that neutrinos might never be revealed. Today we know and we reveal neutrinos produced by different astrophysical objects and by interactions of cosmic rays (natural neutrinos) or produced by nuclear reactors and as secondary particles in accelerators (artificial neutrinos). This paper focus on naturally occurring neutrinos, the disclosure of which requires enormous experimental apparatus in underground laboratories, under water or under the ice of the South Pole. They have allowed huge advances in understanding of neutrino properties with the discovery of the oscillation mechanism. And at the same time they opened new frontiers for the study of astrophysics of the processes that produce energy inside the Sun; on the mechanisms leading to stellar gravitational collapses; on astrophysical objects that produce cosmic rays up to extreme energies.

\end{abstract}
%

\section{Introduzione\label{sec:intro}}

Con la formulazione della struttura teorica (quella che complessivamente oggi chiamiamo \textit{fisica quantistica}) necessaria per la comprensione dei fenomeni atomici e sub-atomici e il raffinamento delle tecniche sperimentali per la rivelazione e la misura delle proprietà  delle particelle (carica elettrica, massa, vita media,...) cominciò, poco meno di un secolo fa, a formarsi il quadro che oggi chiamiamo \textit{modello standard del microcosmo}, di cui il neutrino è un costituente fondamentale.

L'idea dell'esistenza all'interno dei nuclei atomici di una particella neutra e di massa piccolissima (o nulla) fu avanzata nel 1930 da W. Pauli, studiando i dati sperimentali ricavati dai decadimenti radioattivi nucleari. Il fatto che queste particelle non ``preesistessero'' all'interno dei nuclei ma potessero essere ``create e distrutte'' fu una intuizione di pochi anni successiva da parte di Fermi.
Occorrer\`a attendere il 1954, grazie alla nascita dei reattori nucleari e allo sviluppo delle tecniche elettroniche di acquisizione dati per osservarlo.

Questa sintetica rassegna vuole illustrare quale sia il ruolo dei neutrini nel quadro delle nostre conoscenze del microcosmo e del macrocosmo.
Siamo ormai consapevoli che le stelle, oltre la luce visibile, emettono radiazioni di altre lunghezze d'onda che possono essere rivelate con tecniche sperimentali diverse da quelle offerte dai sensori ottici (occhio, telescopio, pellicole fotografiche,...). La possibilità di misurare radiazione nel radio, nell'infrarosso, nell'ultravioletto, nei raggi X e raggi gamma, ha infatti enormemente aumentato la nostra conoscenza dell'universo.

I neutrini sono, come i fotoni, continuamente prodotti da oggetti astrofisici. Come i fotoni, i neutrini possono essere di energia molto bassa (comparabile con quella della radiazione cosmica di fondo) ed arrivare a energie elevatissime: neutrini di energia $>10^6$ volte quella corrispondente all'energia di massa del protone sono stati rivelati. 
Differiscono dai fotoni perché sono estremamente difficili da rivelare, e necessitano di enormi apparati sperimentali, generalmente posti \textit{underground} per essere meglio schermati da altre particelle che penetrano l'atmosfera.

Neutrini sono prodotti in natura dalle stelle durante il loro funzionamento, dalle esplosioni conseguenti il collasso gravitazionale stellare, dalla radioattività presente nei sistemi planetari prodotti dalle esplosioni di supernovae (come succede nel pianeta Terra del sistema Solare). 
Neutrini possono essere prodotti anche artificialmente dall'uomo tramite reattori nucleari per la produzione di energia, e da acceleratori di particelle.
Essendo particelle stabili, i neutrini continuano ad accumularsi e aumentare di numero nell'universo.

In questa rassegna \footnote{Essa è nata a seguito di una presentazione con lo stesso titolo tenuta nell'ambito di un ciclo di eventi in occasione della presenza della mostra dedicata ad Enrico Fermi a Bologna - Ex Chiesa di San Mattia- dal 6 Febbraio al 22 Maggio 2016.} cercherò di illustrare il ruolo dei neutrini prodotti da sorgenti naturali e di energia via via crescente. Tra le cose più rilevanti da segnalare è il fatto che la parola ``neutrino'' non è la traduzione di una parola nata in inglese, o in tedesco. Il \textit{piccolo oggetto neutro} è stato proprio in questo modo battezzato da Enrico Fermi quando iniziò a formulare la teoria che ne descrive le interazioni. 

La rassegna è principalmente rivolta a chi conosce almeno le nozioni di base della fisica classica, e richiama alcuni aspetti di fisica e astrofisica delle particelle. 
Essa è indirizzata in particolare agli insegnanti delle scuole superiori e a quegli studenti che potrebbero avere nella fisica un interesse per il proseguimento degli studi. 
Posso però cercare di sintetizzarla a chi di fisica è digiuno utilizzando la seguente similitudine.
Si pensi ai neutrini come agli spiriti raccontati nel V canto dell'inferno dantesco, il canto dei lussuriosi e di Paolo e Francesca. 
In quel girone, vi è ``un fiato'' che
\vskip 0.2cm
\noindent \textit{di qua, di l\`{a}, di gi\`{u}, di s\`{u} li mena;\\
nulla speranza li conforta mai, \\
non che di posa, ma di minor pena.}
\vskip 0.2cm

\noindent Così, i poveri peccatori carnali sono in eterno, perpetuo movimento. Continuano ad accumularsi e aumentare di numero al passare del tempo. Non hanno contatti tra di loro, e solo occasionalmente due di questi dannati (Francesca e Paolo) interagiscono con il Poeta, che per poterlo fare è dovuto scendere sotto terra. E le informazioni avute dal contatto con una di queste anime ha permesso al poeta di capire il loro stato, e di creare quel capolavoro di poesia.

Così, anche noi ricercatori scendiamo sotto terra, e istalliamo strumentazione nel mare o nel ghiaccio dell'Antartide per incontrare una frazione infinitesima di questi affannati che vagano nell'universo, che possano venire a noi parlar (s'altri nol niega), essere interrogati e raccontarci la storia di come sono stati prodotti e di come sono giunti sino a noi. 


\section{I neutrini nel modello standard del microcosmo\label{sec:storia}}

Allo stato attuale delle nostre conoscenze, le particelle stabili (ossia, che hanno una vita media enormemente più lunga di quella del nostro Universo) sono\footnote{Per maggiori informazioni sulle particelle elementari e sulle loro interazioni si rimanda a \cite{bra11} e alla bibliografia in esso contenuta.}: il fotone \( \gamma  \), i neutrini e gli antineutrini, l'elettrone \( e^{-} \), il positrone \( e^{+} \), il protone \( p \) e l'antiprotone \( \overline p \). 
All'interno dei nuclei che costituiscono la materia ordinaria vi sono anche neutroni, che non possono decadere\footnote{Il decadimento è il processo in cui una particella si trasforma spontaneamente in altre particelle con una vita media caratteristica e rilasciando energia.} quando legati nei nuclei per motivi energetici. 
Neutroni liberi decadono invece con una vita media di circa 15 minuti.
Il protone e il neutrone hanno una struttura interna, essendo costituiti da tre \textit{quark}. 
I quark sono particelle con carica elettrica frazionaria (in termini della carica del protone) che interagiscono tra di loro formando particelle di carica elettrica intera. 
Le interazioni tra quark avvengono attraverso una sorta di \textit{carica forte} (che convenzionalmente chiamiamo \textit{colore}, ma nulla ha a che fare coi colori come convenzionalmente li intendiamo) e sono più complicate delle \textit{interazioni elettromagnetiche} che avvengono tra cariche elettriche. Queste ultime agiscono in modo tale che cariche dello stesso segno si respingano, e cariche opposte si attraggano; l'esistenza di tre \textit{cariche di colore} e tre anticariche di colore rendono le interazioni tra quark molto più complesse.
 
Tutte le particelle composte da quark (ad eccezione del protone) sono instabili e decadono. A livello delle conoscenze attuali, i quark sono oggetti puntiformi e sono quindi davvero oggetti \textit{elementari}.
Altri oggetti elementari sono quelli che chiamiamo \textit{leptoni}, di cui l'elettrone (e la sua antiparticella, il positrone) sono i primi rappresentanti.
Sia i quark che i leptoni si ripetono in tre \textit{famiglie} (o \textit{sapori}). 
Nella prima famiglia di leptoni vi \`{e} l'elettrone ($e^-$ o semplicemente $e$) con il suo neutrino ($\nu_e$); nella seconda il muone ($\mu^-$) e il suo neutrino ($\nu_\mu$); nella terza il tau ($\tau^-$) e il suo neutrino ($\nu_\tau$).
Inoltre, vi sono le famiglie di antiparticelle $(e^+,\overline \nu_e$),
$(\mu^+,\overline \nu_\mu$) e $(\tau^+,\overline \nu_\tau$).
Il modello standard non sa spiegare perché esistano tre famiglie, e lo considera un dato di fatto. Infine, nel modello standard i neutrini erano (sino al 1998) considerate particelle di massa nulla. 
La ragione di questo cambio di prospettiva è uno degli argomenti della seguente storia.

\section{Perché  i neutrini sono difficili da rivelare}
\begin{figure}[tb]
\begin{center}
\includegraphics[width=9.0cm]{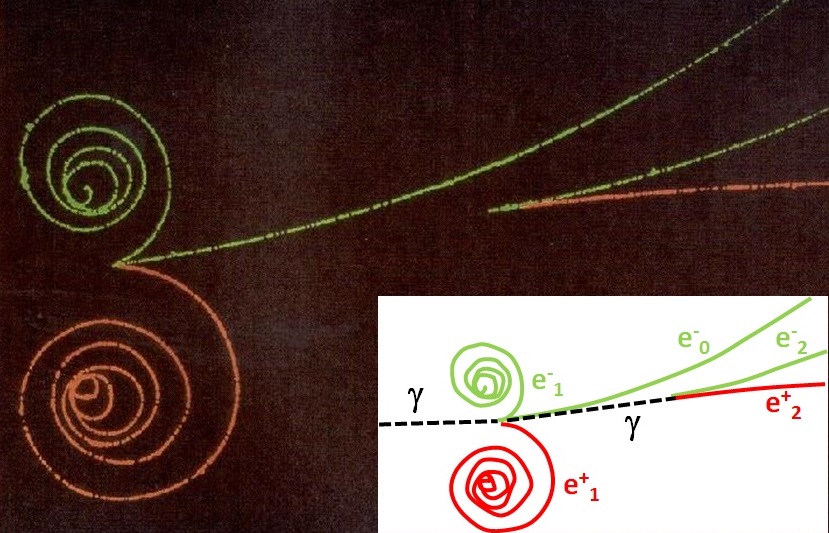}
\end{center}
\caption{\small La rivelazione di un raggio gamma ($\gamma$). Il fotone (privo di carica elettrica) incide in una regione in cui è presente un gas, un campo magnetico e una macchina fotografica. 
 Arrivando da sinistra nella regione con il gas, il fotone urta con un elettrone del mezzo trasferendogli parte della sua energia (traccia verde indicata nel riquadro con $e^-_0$). Parte dell'energia del fotone viene utilizzata per la creazione di una coppia particella-antiparticella (i ricciolini indicati con $e^-_1 , e^+_1$), e il rimanente sotto forma di un fotone di energia più bassa (traccia nera tratteggiata interna alla figura). Questo fotone, infine, sotto l'influenza del campo Coulombiano di un nucleo, converte la sua energia in una seconda coppia (indicata con $e^-_2 , e^+_2$). Le particelle cariche (elettroni e positroni) si rendono visibili perché eccitano e ionizzano il gas, formando dei piccoli centri di addensamento del gas stesso che vengono visualizzati come una minuscola bollicina nella fotografia. La presenza del campo magnetico serve, tramite la forza di Lorentz, per discriminare le tracce positive da quelle negative e per misurare la quantità di moto delle particelle fotografate. Questa grandezza dipende dal raggio di curvature delle particelle visualizzate. }
\label{fig:ep}
\end{figure}

Cosa significa vedere le particelle? Dal punto di vista epistemologico, osservare un oggetto significa rivelare la luce riflessa dalla sua superficie. La luce non è altro che la componente della radiazione elettromagnetica a cui il nostro occhio è sensibile. 
Le dimensioni delle particelle sono tali da risultare enormemente più piccole della lunghezza d'onda della luce visibile (compresa tra 400
nm nel violetto e 700 nm nel rosso), quindi le particelle non possono riflettere la luce. 
L'unica possibilità è dunque quella di \textit{rivelare la radiazione emessa quando le particelle interagiscono con la materia}. Un rivelatore di particelle non è altro che un trasduttore che collega, mediante opportune amplificazioni, un nostro organo di senso o un computer con l'effetto prodotto dall'interazione della particella che si vuole rivelare con il rivelatore stesso. Un esempio storico di come si possono rivelare le particelle - in questo caso, un raggio gamma \footnote{Per motivi storici, fotoni di energia superiore al MeV vengono chiamati \textit{raggi gamma} e indicati con $\gamma$. Vedi poco oltre per il significato di MeV.} - è illustrato in dettaglio in Fig. \ref{fig:ep}.
La fisica delle alte energie è basata su esperimenti in cui le interazioni delle particelle vengono studiate grazie all'uso di rivelatori, più o meno sofisticati.

Il neutrino è una delle particelle stabili che compongono il nostro mondo, ma tra tutte è quella più difficile da rivelare. Il motivo è che esso è privo di carica elettrica, e la sua probabilità d'interazione (e quindi, di produrre particelle con carica elettrica a seguito dell'interazione) è molto piccola.
Il passaggio di particelle cariche in un mezzo provoca infatti processi di \textit{eccitazione} e di \textit{ionizzazione}.
L'eccitazione corrisponde al processo in cui gli elettroni atomici del mezzo attraversato vengono portati a livelli energetici superiori, con conseguente emissione di radiazione elettromagnetica a seguito della diseccitazione. I fotoni di diseccitazione possono essere rivelati da opportuni sensori (chiamati \textit{fotomoltiplicatori, PMT}) che trasducono il segnale ottico in un segnale elettronico.
La ionizzazione corrisponde invece al processo di dissociazione di uno o più elettroni dagli atomi. Gli elettroni liberi possono essere accelerati in presenza di un campo elettrico e produrre un segnale (ad esempio, una scarica su un filo) che può anch'esso essere tradotto nel segnale elettrico di un rivelatore. 

\begin{figure}[tb]
\begin{center}
\includegraphics[width=13.1cm]{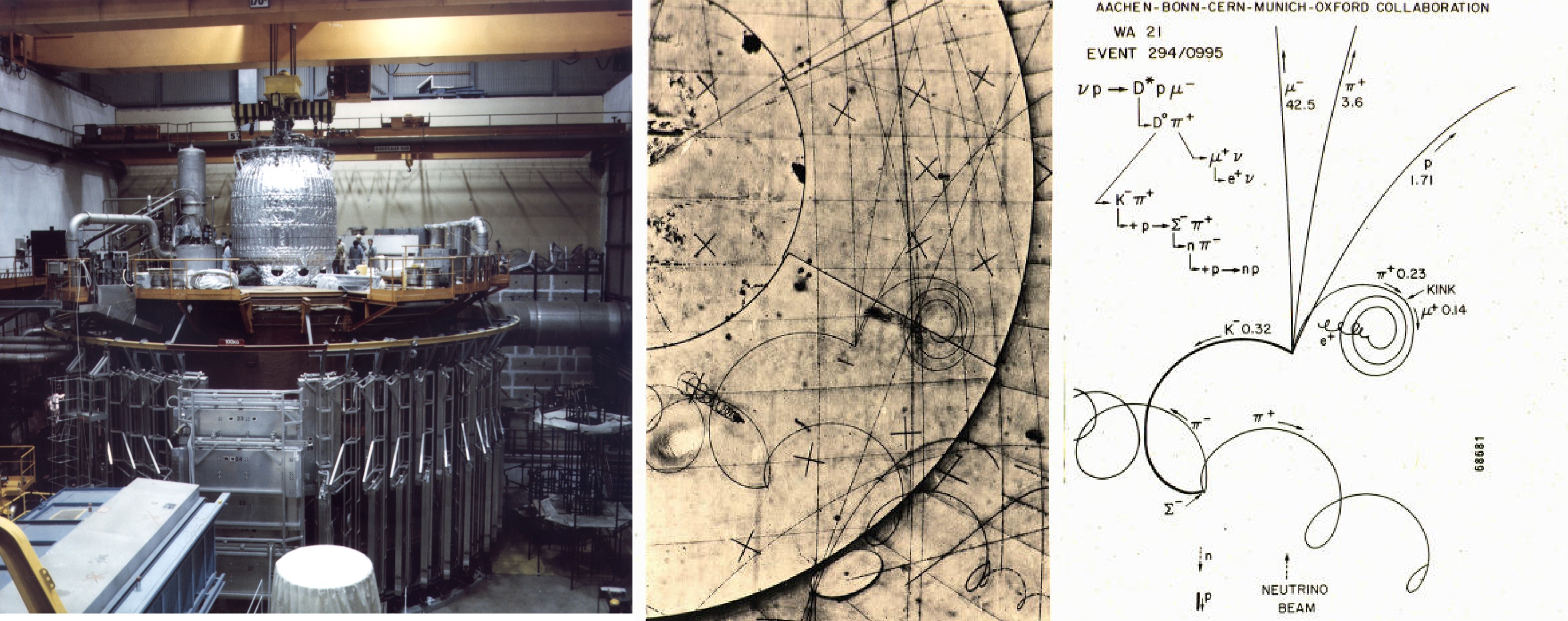}
\end{center}
\caption{\small La grande camera a bolle europea BEBC al CERN (foto a sinistra). Le camere a bolle sono state utilizzate sino agli anni '80 per studiare le particelle, e BEBC in particolare le interazioni dei neutrini. La camera è il cilindro centrale di 3.7 m di diametro. Attorno è equipaggiata con rivelatori elettronici esterni per muoni,  visibili in primo piano. Dietro si trovava la bobina superconduttrice, che produceva un campo magnetico di 24 kgauss. La foto centrale rappresenta una interazione di un $\nu_\mu$ su un protone che produce un $\mu^-$ e una particella instabile composta da quarks ($D^*$), dalla quale discende una serie di decadimenti a catena, di cui i dettagli sono meglio specificati nello schema a destra.}
\label{fig:bubble}
\end{figure}

In analogia col raggio $\gamma$ mostrato nella Fig. \ref{fig:ep}, il neutrino deve interagire per manifestarsi. Interagendo, può ``creare'' particelle cariche a spese della sua energia.
Questo \`e un processo di conversione di energia in massa, attraverso la relazione (universalmente nota) di Einstein: 
\begin{equation}
\label{eq:ein}
E=mc^2
\end{equation}
l'energia a disposizione in un urto fra due particelle si pu\`o trasformare nella massa di nuove particelle, seguendo certe regole ed obbedendo a precise leggi di conservazione.
Tuttavia, al contrario dei raggi $\gamma$ che hanno elevata probabilità d'interazione, i neutrini interagiscono raramente e per rivelarli (come vedremo) sono necessari apparati sperimentali enormi, costosi e difficili da realizzare (vedere Fig. \ref{fig:bubble}).

Un appunto importante sulle unità di misura che utilizzeremo nel seguito. Nel Sistema Internazionale (S.I.) misuriamo le masse in kg, le lunghezze in m, le cariche elettriche in Coulomb (C), le energie in Joule (J), le differenze di potenziale in Volt (V), e così via.
In fisica delle particelle e astrofisica tutto ciò è talvolta scomodo. A tutti è noto che gli astronomi misurano le distanze in anni-luce (ly) o in parsec (pc). In un anno ci sono $3.15\ 10^7$ s; la luce viaggia a $3\ 10^{8}$ m/s; in un anno, la luce percorre quindi $\sim 9.5\ 10^{15}$ m. Un pc corrisponde a $3\ 10^{18}$ cm. Quindi, 1 pc$\simeq $3 ly.

In fisica delle particelle, le energie si misurano in elettronvolt (eV) o multipli (1 MeV e 1 GeV significano rispettivamente $10^{6}$ eV e $10^{9}$ eV).  Un eV è la quantità di energia che un protone (di carica $1.6\ 10^{-19}$ C) guadagna sotto la differenza di potenziale di 1 V. Quindi
\begin{equation}
\label{eq:eV}
 1 \textrm{ eV}= 1.6\ 10^{-19}\ \textrm{C} \times 1\ \textrm{V} = 1.6\ 10^{-19}\ \textrm{CV}= 1.6\ 10^{-19}\ \textrm{J} \ .
\end{equation}
Ma non solo: anche le masse possono essere misurate nelle stesse unità!
Tra massa ed energia esiste la relazione (\ref{eq:ein}). 
Una massa può essere misurata in termini di energia diviso per una costante, la velocità della luce al quadrato, $c^2$.
Quindi posso dire che la massa di un elettrone (per esempio) corrisponde a $m_e= 0.511$ MeV/c$^2$ e quella di un protone a $m_p= 938$ MeV/c$^2$.
Tuttavia, in accordo con la Eq. (\ref{eq:ein}) \textbf{alla massa corrisponde un certo quantitativo di energia}. Quindi se mi domando: quanta energia è sotto forma della massa di un elettrone? La risposta è $m_e c^2 = 0.511 \textrm{MeV/c}^2 \times \textrm{c}^2= 0.511 $ MeV.
E l'energia sotto forma della massa del protone? 938 MeV, ossia quasi 1 GeV.
Molto spesso quindi usiamo, per misurare la massa di una particella, la sua energia a riposo, ossia \textit{la quantità equivalente di energia immagazzinata sotto forma di massa.} Ad esempio, quando un positrone (l'antiparticella dell'elettrone, con carica elettrica opposta e massa esattamente uguale a $m_e=0.511$ MeV/c$^2$) annichila con un elettrone, viene liberata una quantità di energia sotto forma di due fotoni pari a $2\times 0.511$ MeV (è il processo reciproco di quanto mostrato in Fig. \ref{fig:ep}, in cui un fotone crea una coppia $e^-e^+$).
Per avere un riferimento (numeri che è opportuno memorizzare), le transizioni tra livelli atomici producono luce nel visibile, ossia fotoni di energia $\sim 2$ eV. Nelle transizioni nucleari vengono invece emessi fotoni di $\sim 1$ MeV. L'energia a riposo di elettrone e protone corrisponde a circa 0.5 MeV e 1 GeV, rispettivamente.

Quando una particella viene accelerata, oltre all'energia a riposo si aggiunge l'energia cinetica.
Normalmente, le formule classiche (ossia, quelle derivate dalla meccanica di Newton) sono valide fintanto che l'energia cinetica è inferiore all'energia a riposo. Quanto l'energia cinetica supera l'energia di massa, la particella si propaga a velocità ormai prossima a quella della luce e occorre utilizzare le relazioni relativistiche (ossia, quelle introdotte dalla relatività ristretta einsteiniana) per definire grandezze dinamiche, quali energia e quantità di moto.
Per esempio, un elettrone di 100 MeV è una particella relativistica. Un protone di 100 MeV è invece non-relativistico.

Il fotone è una particella di massa nulla (è un \textit{quanto} di energia). Il neutrino, come vedremo nel seguito, è stato ritenuto di massa nulla sino al 1998. Tuttavia, anche se ha massa, essa è così piccola che il neutrino può essere sempre considerato una particella relativistica.

\section{Perché andare in profondità (mare, ghiaccio, terra)? \label{sec:under}}

Nel 1930 si conoscevano soltanto il protone, l'elettrone e il fotone. 
Prima dell'avvento degli acceleratori (verso la fine degli anni '50), la radiazione ionizzante che bombarda costantemente la superficie terrestre, nota sin dal 1912 e chiamata \textit{radiazione cosmica}, contribu\`{i} in modo fondamentale alla comprensione del mondo subatomico. 
L'altissima energia cinetica posseduta dai raggi cosmici (RC) permette infatti la ``creazione'' di nuove particelle. 
{Per un approfondimento dello stato attuale delle conoscenze astrofisiche raggiunte tramite lo studio di particelle cariche (protoni, elettroni e nuclei), antiparticelle, raggi gamma e neutrini si veda \cite{spurio}}.

Le particelle in arrivo sulla sommità dell'atmosfera sono circa 10$^5$ per m$^2$ e per secondo, e vengono chiamati RC \textit{primari}. 
Lo studio dell'origine, dei meccanismi di accelerazione e della propagazione dei RC primari è uno dei più affascinanti campi di studio dell'astrofisica \cite{letetsa}. 
Le particelle prodotte dai primari nell'interazione coi nuclei dell'atmosfera vengono chiamate RC \textit{secondari}.
Le particelle secondarie in arrivo al livello del mare sono ridotte di un fattore circa 1000 rispetto al numero di primari in arrivo sulla sommità dell'atmosfera. 
Inoltre, il passaggio dei RC secondari (principalmente elettroni e muoni) induce effetti biologici molto meno dannosi di quelli che produrrebbero i RC primari (protoni e nuclei). 
Il passaggio di particelle cariche, che produce un segnale nei rivelatori, ha effetti anche sugli organismi. Tutti ormai sanno che la radiazione provoca (tramite il processo di ionizzazione) \textit{radicali liberi}, che portano per esempio all'invecchiamento della pelle. Radicali liberi nei nuclei delle cellule possono produrre malfunzionamenti delle stesse, innescando tumori.
Ecco un altro motivo del perché la presenza di atmosfera ha contribuito allo sviluppo della vita.

Per ridurre ulteriormente il flusso di particelle ionizzanti in un rivelatore occorre andare in profondità. L'acqua, il ghiaccio oppure il terreno e la roccia terrestre provvedono a ridurre ulteriormente il fondo dovuto ai RC secondari.
Ad esempio, nei Laboratori del Gran Sasso (accessibile dalla galleria autostradale che connette Roma col mare Adriatico) il flusso di RC secondari (particelle cariche) è ridotto di circa un milione di volte rispetto a quella misurabile al livello del mare.
Per questo motivo, tutti gli esperimenti che richiedono un bassissimo fondo di radiazione ambientale sono disposti in profondità (\textit{underground}). I \textit{telescopi di neutrini}, che descriveremo in \S \ref{sec:neutel}, operano sotto grandi spessori di acqua o ghiaccio.

Lo sviluppo di laboratori {underground} partì dalla fine degli anni '70-inizio anni '80, con la nascita di alcune teorie che prevedevano che il protone non fosse stabile. Per la verifica sperimentale, cominciarono a nascere esperimenti che dovevano tenere sotto osservazione una massa dell'ordine di qualche migliaia di tonnellate, e che dovevano funzionare il più possibile al riparo di RC secondari. 
Apparati sperimentali vennero quindi installati in prossimità di miniere in disuso o, meglio, in cavità di servizio presso tunnel autostradali. Queste ultime erano ovviamente più semplici da raggiungere (spesso gli esperimenti in miniere sono raggiungibili solo con cigolanti montacarichi). Le sale sperimentali del Gran Sasso vennero appositamente realizzate all'epoca della costruzione dell'autostrada Roma-Teramo.

\begin{figure}[tb]
\begin{center}
\includegraphics[width=10.5cm]{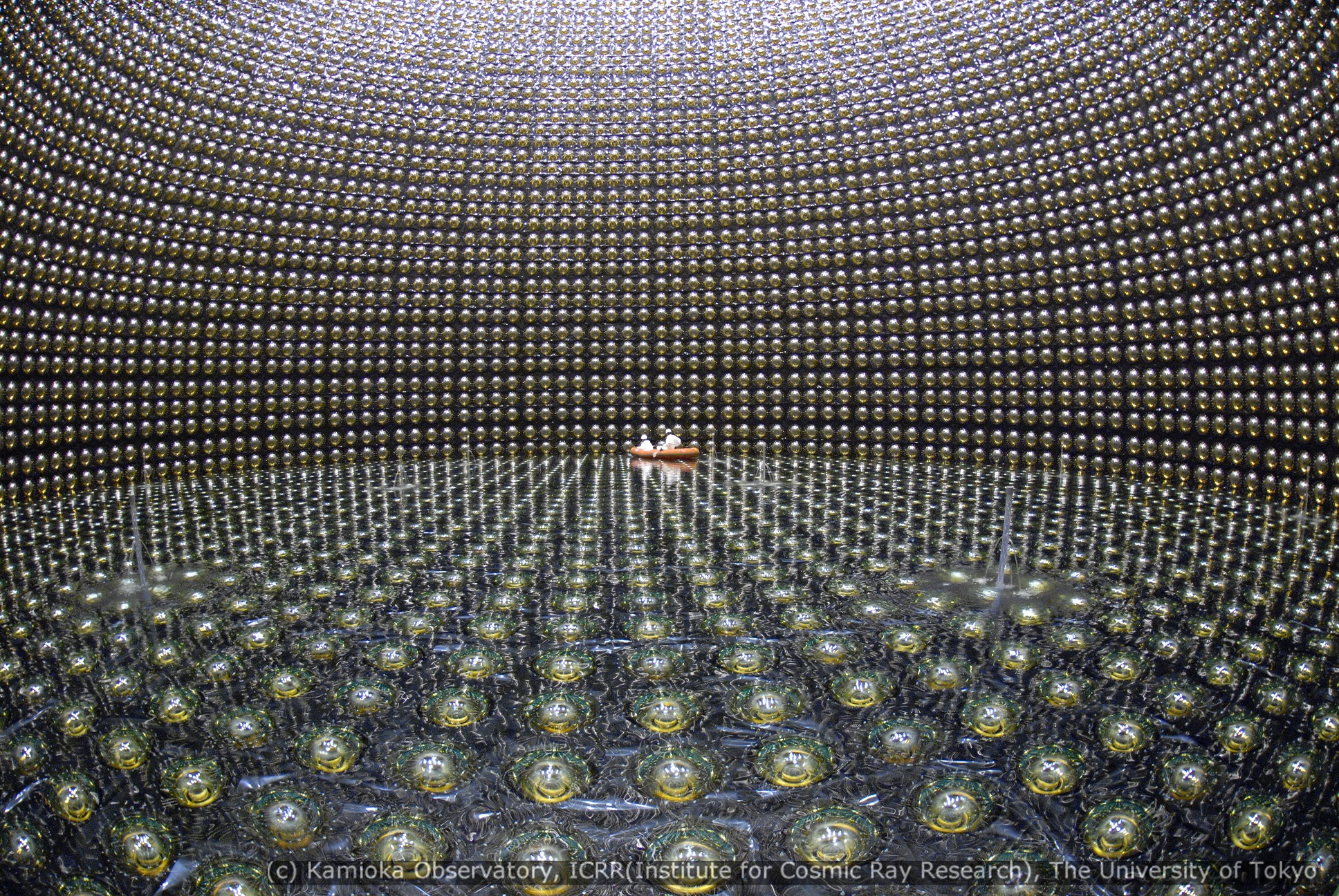}
\end{center}
\caption{\small Fotografia dell'esperimento Super-Kamiokande in Giappone, in funzione dal 1996. È un cilindro di 39 m di diametro e 42 m di altezza riempito di acqua purificata. Circa 11000 fotomoltiplicatori (PMT) di quasi 70 cm di diametro ricoprono circa il 40\% della superficie interna del cilindro. Quando una particella carica attraversa l'acqua, viene indotta luce Cherenkov; la luce, giungendo sui PMT produce un segnale elettrico proporzionale alla sua intensità. Analizzando i PMT interessati dallo stesso evento, è possibile ricostruire se la particella è un elettrone o un muone (quindi, prodotta da $\nu_e$ oppure $\nu_\mu$), la sua direzione e la sua energia.
La foto è stata scattata durante la fase di riempimento, con dei tecnici che controllano i PMT per mezzo di un canotto. Crediti: Kamioka Observatory, ICRR (Institute for Cosmic Ray Research), The University of Tokyo.}
\label{fig:sk}
\end{figure}

Schermati dai RC secondari carichi, questi esperimenti cominciarono a funzionare ma si accorsero di un altro fondo, questo irriducibile anche con l'aumentare della profondità: quello dovuto ai \textit{neutrini atmosferici}. I neutrini, prodotti anch'essi dalle interazioni dei RC primari coi nuclei dell'atmosfera, non vengono significativamente ridotti dalla presenza dell'atmosfera o di roccia (o acqua, ghiaccio) sovrastante. Tuttavia, occasionalmente, i neutrini possono interagire nel rivelatore e produrre particelle cariche che potranno produrre un segnale. 
Quello dei neutrini atmosferici, che inizialmente sembrava un fastidioso disturbo, ha prodotto un risultato di prima importanza in fisica delle particelle (riconosciuto col Nobel 2015 a T. Kajita per l'esperimento Super-Kamiokande, SK). Infatti nel 1998 l'esperimento SK situato in una miniera in Giappone (Fig. \ref{fig:sk}) riportò una discrepanza tra il numero di neutrino misurati e attesi. La stessa discrepanza, seppure con minore significatività statistica, venne contemporaneamente osservata dall'esperimento MACRO situato in una sala sperimentale del Gran Sasso.
Questa discrepanza verrà interpretata come un effetto del fatto che la massa dei neutrini non è nulla (come invece precedentemente ritenuto).
Tratteremo di questo nel \S \ref{sec:osci}.

Tutte le osservazioni di neutrini che discuteremo nel seguito sono avvenute (e avvengono) in esperimenti situati sotto chilometri di roccia, acqua o ghiaccio.

\section{La storia dei neutrini (e non solo)\label{sec:sto1}}

Il processo radioattivo dei nuclei che comunemente chiamiamo \textit{decadimento beta} (vedere Cap. 14 di \cite{bra11}) rappresenta una  trasmutazione di un elemento $(Z,N)$, ove $Z$ \`{e} il numero di protoni ed $N$ quello di neutroni, verso un nucleo ($Z+1,N-1$)  (decadimento \textit{beta negativo}), oppure ($Z-1,N+1$) (decadimento \textit{beta positivo}). 
Era noto sin dall'inizio del 1900 che nel caso di transizioni \textit{beta negative}:
\begin{equation}
\label{eq:8-in1}
(Z,N) \rightarrow (Z+1, N-1) + e^-
\end{equation}
un elettrone veniva emesso dal nucleo. L'energia posseduta dall'elettrone era tipicamente di parecchi MeV, molto maggiore dell'energia a riposo dell'elettrone (0.511 MeV).
\begin{figure}[tb]
\begin{center}
\includegraphics[width=7.5cm]{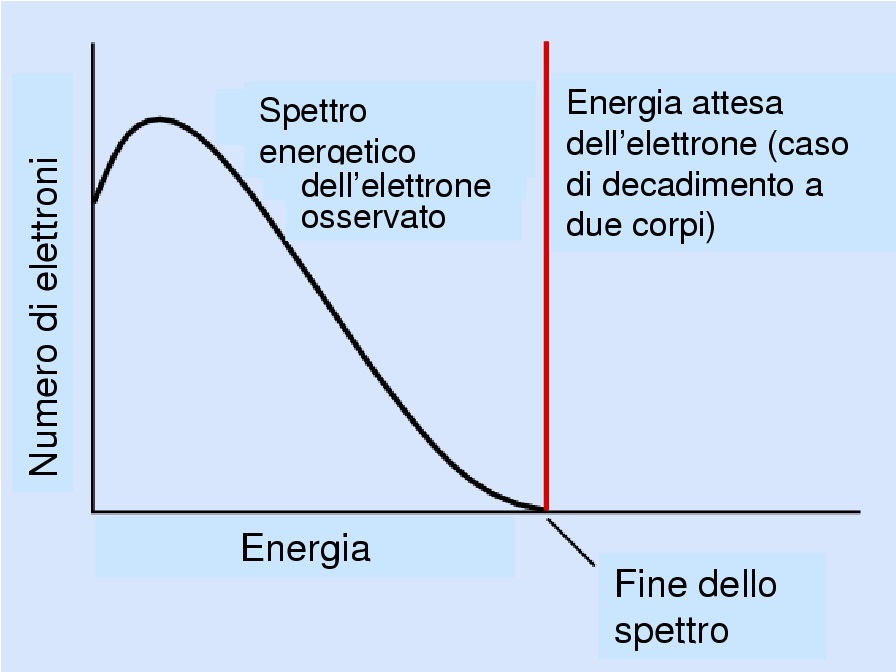}
\end{center}
\caption{\small Forma della distribuzione dell'energia trasportata dall'elettrone nel decadimento beta di un nucleo. Lo spettro atteso nel caso di decadimento a due corpi coinciderebbe con una riga, al valore che corrisponde alla fine dello spettro della curva continua misurata.}
\label{fig:beta}
\end{figure}

Se un nucleo a riposo subisce il decadimento (\ref{eq:8-in1}), la conservazione dell'energia e dell'impulso impongono che il nucleo prodotto e l'elettrone rinculino nella stessa direzione e verso opposto. Tuttavia, poiché il nucleo ha massa almeno migliaia di volte maggiore di quella dell'elettrone, la sua velocit\`{a} di rinculo \`{e} trascurabile. 
Quindi l'elettrone deve essere emesso con energia costante (valore evidenziato in Fig. \ref{fig:beta} dalla linea parallela all'asse delle ordinate), coincidente in pratica con tutta l'energia rilasciata nel decadimento.

Tuttavia, i risultati sperimentali (usando qualsivoglia nucleo) erano in completo disaccordo con quanto sopra. 
Infatti, l'elettrone possedeva uno spettro continuo di energia, sino a raggiungere il valore massimo previsto (ossia, quello corrispondente al fatto che l'elettrone trasporti tutta l'energia), come presentato dalla curva continua di Fig. \ref{fig:beta}.
In pratica, il decadimento (\ref{eq:8-in1}) \textbf{sembrava violare} la legge di conservazione dell'energia (e del momento angolare). 

Questo portò nel 1930 W. Pauli a formulare l'ipotesi dell'esistenza nella reazione (\ref{eq:8-in1}) di una particella aggiuntiva preesistente nel nucleo:
\begin{equation}
\label{eq:8-in2}
(Z,N) \rightarrow (Z+1 , N-1) + e^- + \overline \nu_e \ .
\end{equation}
Questo nuovo oggetto emesso doveva essere privo di carica elettrica e doveva trasportare l'energia, la quantità di moto e il momento angolare che risultavano mancanti nelle osservazioni sperimentali
\footnote{Nella equazione (\ref{eq:8-in2}) con il simbolo $\overline \nu_e$ stiamo aggiungendo informazioni aggiuntive rispetto quanto ipotizzato da Pauli, in particolare che nel decadimento beta negativo viene sempre emesso un antineutrino elettronico (indicato appunto $\overline \nu_e$). Viceversa, nel decadimento beta positivo viene sempre emesso un neutrino elettronico (indicato con $\nu_e$). Questi, compongono quella che si chiama la prima famiglia di neutrini. 
Altre due famiglie, quella del neutrino muonico e del neutrino tauonico, verranno scoperte successivamente.}. 
Poiché un tale oggetto non era mai stato osservato, Pauli assunse che fosse altamente elusivo, e non soggetto alle interazioni allora conosciute.
La preoccupazione di Pauli era come rivelare tale particella. Egli temeva che la verifica sperimentale della sua ipotesi non fosse assolutamente realizzabile. 

Fu Enrico Fermi a battezzare nel 1934 la particella di Pauli con il nome \textit{neutrino}. Non solo, egli iniziò la formulazione della teoria matematica che permette di descrivere il decadimento beta in \textit{termini quantitativi}\footnote{Ciò significa che la teoria, per ogni nucleo, permette di predire se il nucleo è stabile, o quanto tempo in media vive prima di trasmutarsi in altro nucleo.}, teoria poi sostanzialmente passata nella successiva formulazione delle \textit{interazioni deboli}. 
Uno dei meriti ulteriori della teoria di Fermi fu quello di suggerire la possibilit\`{a} di una reazione in cui i neutrini potevano interagire con la materia (ossia, con protoni, neutroni ed elettroni).
Tuttavia ancora nel 1936 Bethe e Bacher affermavano: \textit{sembra praticamente impossibile rivelare neutrini liberi, ossia dopo che sono stati emessi dall'atomo radioattivo. Esiste una sola reazione che neutrini possono causare: il processo $\beta$ inverso, cio\`{e} la cattura di un neutrino da parte di un nucleo, accompagnata con l'emissione di un elettrone (o positrone)}.
Ciò corrisponde al processo:
\begin{equation}
\overline \nu_{e} + p\rightarrow e^{+} + n \ .
\label{eq:8-nu1}
\end{equation}
che sarebbe stato effettivamente usato, quasi 20 anni dopo, per la scoperta. 

Da lì a poco, la II guerra mondiale avrebbe provocato cataclismi in tutto il mondo, e anche nella fisica. 
Fermi, insignito del premio Nobel nel 1938, preso il premio da Stoccolma si imbarc\`{o} diretto negli Stati Uniti fuggendo dall'Italia: Mussolini stava emanando le infami leggi razziali e la moglie di Fermi era ebrea. 
Molti fisici europei rifugiati negli Stati Uniti, insieme a Fermi, ebbero un ruolo rilevante nel progetto Manhattan: quello che avrebbe portato alla realizzazione delle  {bombe nucleari} e al successivo sviluppo di reattori nucleari per la produzione di energia per fini pacifici. 
Proprio uno di questi reattori, sorgente di un abbondante flusso di neutrini, giocò un ruolo fondamentale per la loro scoperta. 

Come suggerito da Bethe e Bacher, dalla teoria di Fermi era chiaro che la possibilit\`{a} di rivelare sperimentalmente i neutrini era legata alla reazione (\ref{eq:8-nu1}).
Una ulteriore difficoltà era legata alla necessità di discriminare l'interazione di un $\overline \nu_{e}$ da un comune evento di decadimento $\beta$ positivo di un nucleo. 
Qualunque rivelatore ha una piccolissima contaminazione di materiale radioattivo i cui decadimenti costituiscono un \textit{fondo irriducibile} per la reazione cercata. 
Esistevano quindi due ordini di problemi per poter essere certi che in processo (\ref{eq:8-nu1}) avesse avuto luogo:
1) la probabilit\`{a} d'interazione del neutrino prevista dalla teoria di Fermi \`{e} cos\`{i} piccola che occorre un flusso enorme di $\overline \nu_{e}$  perché si abbia la possibilit\`{a} di osservare qualche interazione.
2) Nello stato finale della (\ref{eq:8-nu1}) compare un positrone e un neutrone; quest'ultima \`{e} una particella neutra non banale da rivelare. Tuttavia, il neutrone $deve$ essere osservato, per essere sicuri che il positrone misurato non sia quello proveniente da un decadimento di un nucleo radioattivo.

\begin{figure}[tb]
\begin{center}
\includegraphics[width=9.5cm]{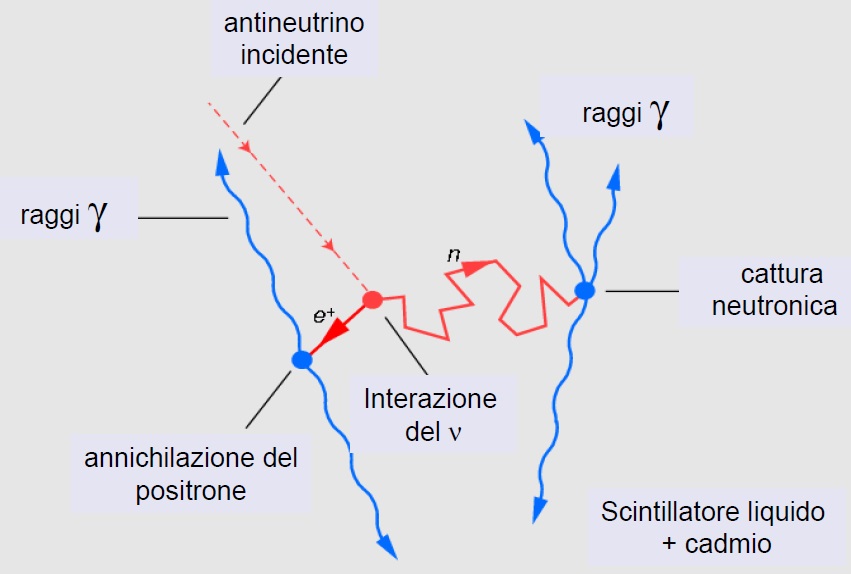}
\end{center}
\caption{\small Schema della misura dei prodotti dovuti alla reazione (\ref{eq:8-nu1}). Vi \`{e} un segnale immediato dovuto ai raggi $\gamma$ indotti dall'annichilazione di un positrone con un elettrone del mezzo. Il neutrone viene prima rallentato dall'acqua e poi catturato, producendo altri raggi $\gamma$ ritardati di un tempo sufficientemente lungo da produrre un secondo segnale nell'elettronica di acquisizione. }
\label{fig:nudisc}
\end{figure}

Nel 1954 negli USA i fisici C. Cowan e F. Reines ebbero l'idea di utilizzare come sorgente di neutrini un reattore nucleare della potenza di circa 150 Megawatt. Vi era la possibilit\`{a} d'istallare un rivelatore a circa 11 metri dal $core$ del reattore, e a circa 12 m di profondit\`{a}. A quella distanza, quando il reattore era al massimo della potenza, il numero di $\overline \nu_e$ in arrivo dal $core$ era di $ \simeq 10^{13}\ cm^{-2}s^{-1}\ $. 
Il rivelatore consisteva di due contenitori di 200 litri di acqua\footnote{Il numero di protoni bersaglio in 200 kg di acqua corrisponde a $0.6\times 10^{28}$ protoni.}, disposti tra tre contenitori di scintillatore liquido, ciascuno della capacit\`{a} di 1400 litri. Nell'acqua (e questo era un punto geniale dell'esperimento) era disciolto un sale di cadmio, elemento che ha una elevata affinità per catturare neutroni liberi.
A seguito della cattura di un neutrone, il nucleo di cadmio si trova in uno stato eccitato che ritorna allo stato fondamentale con l'emissione di raggi $\gamma$ per circa 8 MeV. Questo secondo segnale permette di ``osservare'' la presenza del neutrone nell'evento, Fig. \ref{fig:nudisc}.
L'esperimento era minuscolo sulla scala degli esperimenti attuali, che descriveremo nelle successive sezioni.
I ricercatori effettuarono numerose prove per assicurarsi che il segnale era autenticamente dovuto alla reazione dell'antineutrino anziché ad altre reazioni di fondo. 

L'esperimento prese dati per 5 mesi: 900 ore con il reattore alla massima potenza, e 250 ore con il reattore spento. L'elettronica di acquisizione era tale da poter acquisire il segnale dovuto all'annichilazione del positrone, e quello indotto dai $\gamma$ emessi dal cadmio dopo la cattura del neutrone, ritardato di qualche $\mu s$ (l'occorrenza di entrambi i segnali è chiamata \textit{coincidenza ritardata}). 
Qualitativamente, si vide immediatamente che il numero di eventi in coincidenza ritardata erano molto pi\`{u} numerosi (4 o 5 volte maggiori) col reattore $on$ anziché col reattore $off$. 
L'esperimento permise non solo di affermare l'esistenza del neutrino (in termini moderni, esattamente dell'$\overline \nu_e$), ma permise anche di verificare quantitativamente la teoria del neutrino di Fermi.

Pochi anni dopo questa scoperta, venne identificato il neutrino della seconda famiglia, ossia il neutrino muonico $\nu_\mu$.
Mentre i neutrini della prima famiglia quando interagiscono producono sempre elettroni (e gli antineutrini $\overline \nu_e$ producono sempre positroni $e^+$), i neutrini della seconda famiglia quando interagiscono producono sempre $\mu^-$ (e gli $\overline \nu_\mu$ sempre $\mu^+$).
I muoni si distinguono dall'elettrone perché hanno massa maggiore e sperimentalmente lasciano segnali differenti nei rivelatori di particelle.
Il segno della carica elettrica pu\`{o} essere discriminato facendo passare le particelle in una regione con campo magnetico (che deflette da una parte le particelle negative, e nel verso opposto quelle positive).
Infine, nel 2000 venne rivelata la prima interazioni dei neutrini tauonici nell'esperimento DONUT negli USA.

Da informazioni che abbiamo dal Large Electron Positron (LEP) collider al CERN, oggi sappiamo che il quadro delle tre famiglie di neutrini è completo.

\section{{}``Vita''. Ossia, i neutrini e il funzionamento delle stelle\label{sec:vita}}

Una volta che si realizzò che i neutrini potevano essere rivelati, si immaginò che potessero essere utilizzati per risolvere uno dei più importanti problemi filosofici dell'uomo: cosa sono  e come funzionano le stelle\footnote{Nelle presentazioni divulgative che tengo sull'argomento, mostro talvolta un breve estratto dal film \textit{Il Re Leone} della Walt Disney (reperibile su \url{https://www.youtube.com/watch?v=HTn7oDidayA}). Simba (il futuro re) e i suoi amici Timon (T) e Pumbaa (P) sono in quiete dopo essersi nutriti, ed osservano il cielo di una notte stellata. Qui avviene il seguente scambio di battute: 
\textbf{P: }\textit{Timon, ti sei mai domandato che cosa siano quei lumicini lassù?} 
\textbf{T: }\textit{Pumbaa, io non mi faccio domande. Io le cose le so!... Sono delle lucciole, lucciole che sono rimaste attaccate a quell'enorme cosa nero-bluastra...}
\textbf{P: }\textit{Oh sì! Io pensavo che fossero masse gassose che bruciano a miliardi di km di distanza!}. 
Come questi personaggi fanno riflettere già ai bambini, filosofi e pensatori sin dalla remota antichità hanno cercato la risposta alla domanda di Pumbaa. Qualcuno \textit{ipse dixit} aveva, come Timon, la risposta. Ma solo i neutrini ci hanno dimostrato che avevamo finalmente risolto il problema.}.

Nel corso del 1800, si iniziarono a utilizzare teorie fisiche per spiegare il funzionamento del Sole e tentare di derivarne l'età. Assumendo una combustione chimica della massa solare, si derivava un'età di poche migliaia di anni (compatibili con un modello \textit{creazionistico} dell'universo). 
William Thomson (Lord Kelvin), immaginando che la sorgente di radiazione fosse dovuta alla trasformazione di energia di legame gravitazionale, stimò un'età del sistema solare dell'ordine dei milioni di anni. 
Darwin, che nella sua teoria evoluzionistica prevedeva scale temporali molto più lunghe, aveva molta stima di Lord Kelvin e per questo tolse ogni riferimento a possibili scale temporali nelle edizioni del suo ``L'origine delle specie'' successive alle predizioni di Thomson.
Inoltre affermò che la questione dell'età della Terra fosse ``uno dei miei guai più gravi''.

La persona che per primo immaginò che la sorgente di energia nel Sole fossero reazioni di fusione nucleare fu Hans Bethe verso la fine degli anni '30. Poiché ogni reazione nucleare produce $\sim 10^6$ più energia di una reazione di tipo atomico, la vita del Sole aumenta di un milione di volte rispetto a quanto determinato stimando che l'energia sia prodotta da reazioni chimiche. L'età del Sole e del suo sistema passa da qualche migliaio di anni a diversi miliardi di anni; questo avrebbe completamente tranquillizzato il povero Darwin. 
Finalmente, a partire dagli anni '60, si iniziò a pensare come \textit{verificare sperimentalmente} questo modello. Si intuì (principalmente da parte di John Bahcall \cite{bahcall}) che la chiave di tutto stava nella possibilità di rivelare quei neutrini che, inevitabilmente, debbono essere prodotti dalle reazioni di fusione nucleare, se esse avvengono nel Sole.

Nella nostra comprensione attuale, gli equilibri reciproci tra interazioni gravitazionali, elettromagnetiche e nucleari determinano l'evoluzione delle stelle.
La pressione gravitazionale comprime i nuclei nel centro della stella, avvicinandoli tra di loro contro la repulsione Coulombiana che agisce tra cariche dello stesso segno. L'enorme pressione forma un plasma di densit\`{a} cos\`{i} elevata da innescare delle reazioni di fusione tra i nuclei pi\`{u} leggeri. Le reazioni di fusione producono energia sotto forma di fotoni dell'ordine del MeV.
Una stella \`{e} un sistema in equilibrio tra la pressione dovuta alla gravit\`{a} e la pressione verso l'esterno dovuta alla radiazione prodotta dalle reazioni di fusione nel nucleo.

\begin{figure}[tb]
\begin{center}
\includegraphics[width=13.5cm]{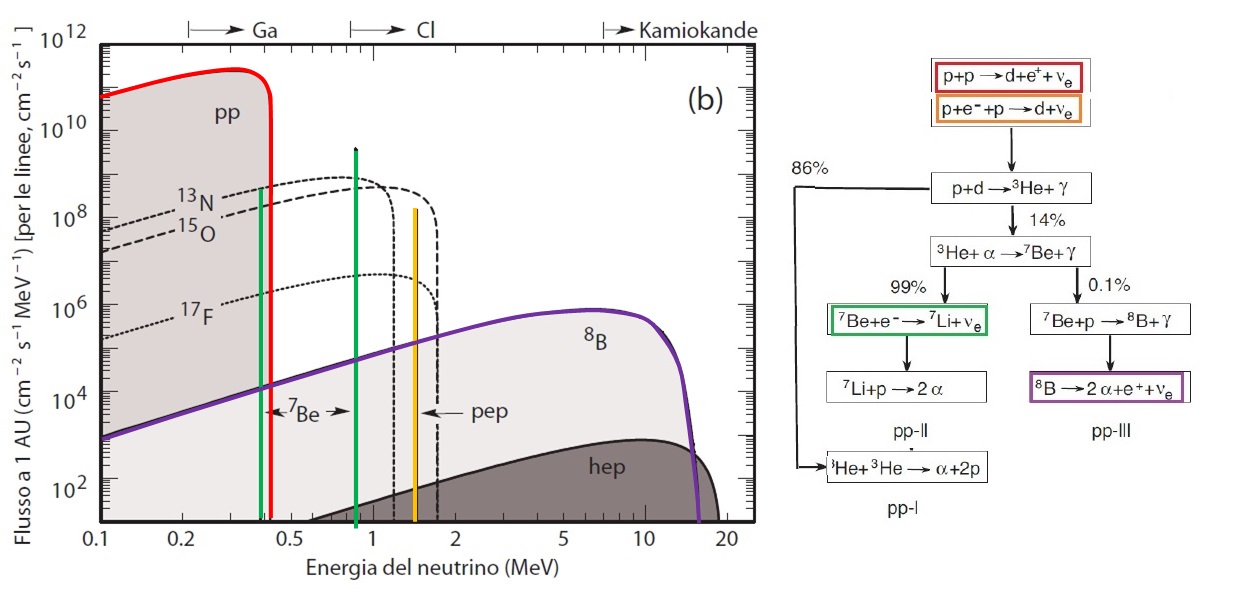}
\end{center}
\caption{\small Spettro energetico dei neutrini solari in arrivo sulla Terra: le linee intere indicano i neutrini provenienti dalle reazioni del ciclo pi\`{u} importante (ciclo \protect\( pp\protect \)), quelle tratteggiate indicano i neutrini provenienti dal ciclo CNO. A destra: catena delle principali reazioni nucleari che avvengono al centro del Sole. Quelle che producono neutrini sono evidenziate con gli stessi colori riportati nella parte sinistra della figura.  }
\label{fig:sole}
\end{figure}

Il Sole fornisce un'opportunit\`{a} unica per verificare la teoria di come le stelle funzionino, perché è la stella più vicina a noi.
Il ``reattore termonucleare'' centrale \`{e} molto pi\`{u} piccolo delle dimensioni solari; le reazioni di fusione che vi avvengono sono sintetizzabili come:
\begin{equation}
\label{eq:12-53}
4p \rightarrow ^4\textrm{He} + 2e^+ +2\nu_e + Q \quad; \quad\quad Q=26.73\ \textrm{MeV} 
\end{equation}
Ossia: 4 protoni, a seguito di diversi processi riportati in dettaglio nella Fig. \ref{fig:sole}, si fondono per formare un nucleo di elio ($^4\textrm{He}$). Nel processo viene complessivamente emessa energia sotto forma di fotoni pari a Q=26.73 MeV. Per ogni nucleo di elio formato\footnote{Il nucleo di elio \`{e} composto da 2 protoni e 2 neutroni. Quindi, due dei protoni iniziali nella (\ref{eq:12-53}) hanno subito il decadimento $p\rightarrow n + e^+ + \nu_e$.}, vengono emessi due $e^+$ (che annichilano con due elettroni solari) e due $\nu_e$, che invece fuggono immediatamente, a seguito della loro piccola probabilità d'interazione.
Al contrario dei $\nu_e$, i fotoni (che hanno processi elettromagnetici con la materia) subiscono un gran numero di collisioni e impiegano di fatto un lunghissimo tempo per giungere alla superficie del Sole (\( \sim 0.2 \) milioni di anni). 
Quando arrivano sulla superficie solare, la fotosfera, questi sono ridotti alle lunghezze d'onda e alle energie caratteristiche con cui il Sole ci appare visibile. 

Poiché la luminosit\`{a} solare (cio\`{e} quanta energia viene emessa dalla fotosfera solare al secondo) \`{e} nota e pari a $L_\odot= 3.842\ 10^{26}$ W, sapendo che 1 MeV=$1.6\ 10^{-13}$ J, il numero di neutrini per secondo attesi sulla Terra (che dista mediamente $D_\odot=  1.495\ 10^{13}$ cm dal Sole) \`{e} dato dalla seguente relazione:
\begin{equation}
\label{eq:12-S1}
\Phi_{\nu_e} \simeq {1\over 4\pi D^2_\odot} {2L_\odot\over Q} \simeq 6\times 10^{10} \textrm{  cm}^{-2}\textrm{s}^{-1} \ .
\end{equation}
Questa relazione ci dice una cosa estremamente semplice: attraverso una superficie delle dimensioni all'incirca del polpastrello del nostro pollice, passano 60 miliardi di neutrini per secondo originati nel Sole!
Dalla massa e della luminosit\`{a} si può dedurre che una stella come il Sole può funzionare per circa 10 miliardi di anni. 
Il Sole ha convertito H in He per circa $ 4.5 \times 10 ^ 9 $ anni e il processo continuer\`{a} per circa altrettanti anni.

Il modello che spiega il comportamento del Sole non solo predice il numero totale di neutrini, ma anche il loro \textit{spettro energetico}.
 Lo spettro energetico corrisponde al numero di neutrini emessi in funzione della loro energia, ed \`{e} mostrato nella parte sinistra di Fig. \ref{fig:sole}. I dettagli delle reazioni che avvengono sono invece nella parte destra della figura.
La maggior parte dei neutrini emessi proviene dalla reazione \( pp\rightarrow de^{+}\nu _{e} \), che produce neutrini con energie comprese fra 0 e 0.42 MeV. I pochi neutrini di maggior energia (fino a 14.06 MeV) provengono dal decadimento del \( ^{8} \)B. Vi sono anche neutrini monocromatici, per esempio quelli dovuti al decadimento del \( ^{7} \)Be. 

\subsection{Esperimenti per la rivelazione di neutrini dal sole\label{sec:vitaexp}}

Gli esperimenti per la misura del flusso di \( \nu _{e} \) solari che investe la Terra sono estremamente complessi.
 
Il primo esperimento, ideato da R. Davis (Nobel nel 2002) su sollecitazione e stimolo di J. Bahcall, fu quasi temerario per le difficoltà che doveva affrontare. Dopo una lunga fase di progettazione, il rivelatore inizi\`{o} a prendere dati all'inizio degli anni '70 nella miniera di Homestake negli USA. 
Si trattava di un esperimento radiochimico che usava come bersaglio un grande rivelatore contenente una soluzione di cloro, dove avveniva la reazione 
\begin{equation}\label{eq:Cl}
\nu _{e}+^{37}\mbox {Cl}\rightarrow ^{37}\mbox {Ar}+e^{-} \ .
\end{equation} 
Questa reazione ha una soglia energetica di 0.814 MeV; quindi solo i neutrini provenienti dal decadimento del \( ^{8} \)B e dalla cattura elettronica nel \( ^{7} \)Be possono essere ri\-ve\-la\-ti (vedi Fig. \ref{fig:sole}).
I risultati sperimentali con il \( ^{37} \)Cl indicavano un flusso
di neutrini \( \nu _{e} \) pari a un terzo di quelli predetti dal modello standard del Sole. Con questo risultato inizi\`{o} il problema dei neutrini solari. 
Assumendo corretto l'esperimento (e molti inizialmente non credevano a questa possibilit\`{a}) due casi erano dati: 
il modello solare sovrastimava la produzione di neutrini (molti non credevano alla correttezza delle predizioni di J. Bahcall), e doveva essere perfezionato;
oppure che, tra il centro del Sole e la Terra, qualcosa accadeva al neutrino, ad esempio le \textit{oscillazioni} dei \( \nu _{e} \) in neutrini di diverso sapore (\( \nu _{\mu } \), \( \nu _{\tau } \)) non osservabili negli esperimenti. 
Col passare degli anni, \`{e} risultato che l'esperimento non sbagliava,  il modello del Sole \`{e} corretto, e che i neutrini oscillano (vedi \S \ref{sec:osci}).

All'inizio degli anni '90 entrarono in funzione due altri esperimenti radiochimici (Gallex, poi GNO, al Gran Sasso e Sage in Russia) che utilizzavano come bersaglio il \( ^{71} \)Ga, ed erano sensibili a neutrini con energia superiore a 0.233 MeV, tramite il processo 
\begin{equation}\label{eq:Ga}
\nu _{e}+^{71}\textrm{Ga}\rightarrow ^{71}\textrm{Ge}+e^{-} \ .
\end{equation} 
La rivelazione dei neutrini solari con \( E_{\nu }>0.233 \) MeV (le soglie energetiche in cui gli esperimenti sono sensibili sono riportate in alto in Fig. \ref{fig:sole}) include i neutrini prodotti nella reazione $p+p\rightarrow d+ e^+ + \nu_e$ e ha dimostrato che effettivamente il Sole ha al suo centro una {}``centrale a fusione nucleare''. Anche questi esperimenti radiochimici che utilizzavano gallio hanno riportato un significativo deficit di neutrini solari.

Altri due esperimenti, che si basano su un meccanismo di rivelazione differente, confermarono il deficit di $\nu_e$ dal Sole. 
I due esperimenti (Kamiokande e Super-Kamiokande, vedi Fig. \ref{fig:sk}, in Giappone) hanno funzionato in sequenza, essendo il secondo concettualmente identico al primo, ma realizzato su una scala più grande.
Qui il bersaglio è fornito dell'acqua contenuta in enormi contenitori.
I neutrini provenienti dal Sole interagiscono tramite collisioni elastiche sugli elettroni:\footnote{Nota per i pi\`{u} esperti. In realt\`{a} il processo di ES pu\`{o} essere prodotto anche da $\nu_\mu$ e $\nu_\tau$ attraverso il \textit{processo a corrente neutra}. Questo per\`{o} comporta una probabilit\`{a} d'interazione per $\nu_\mu$, $\nu_\tau$ molto pi\`{u} piccola che per l'interazione a corrente carica del $\nu_e$.}
\begin{equation}\label{eq:ES}
ES:\quad\quad \nu _{e}+e^{-}\rightarrow \nu _{e}+e^{-}\ .
\end{equation} 
Come in ogni processo elastico, la particella urtata (in questo caso uno degli elettroni dell'acqua) si muove {}``ricordando'' la direzione di provenienza del neutrino, ossia la posizione del Sole\footnote{In realtà il neutrino è partito nella posizione del cielo dove si trovava il Sole 8.5 minuti prima.}.
Le interazioni dei neutrini solari avvengono sia di giorno che di notte, poiché la Terra praticamente non ha alcun potere di assorbimento per neutrini di queste energie\footnote{La probabilità d'interazione dei neutrini aumenta con l'aumentare dell'energia. I neutrini cominciano ad essere assorbiti nell'attraversare il diametro terrestre ad energia di $\sim 10^{14}$ eV, ossia ad energie cento milioni di volte maggiori di quelle dei neutrini solari.}.
Questi esperimenti sono sensibili ai neutrini solari di più alta energia, ossia quelli provenienti dal processo \( ^{8} \)B (Fig. \ref{fig:sole}). 

\begin{figure}[tb]
\begin{center}
\includegraphics[width=12.5cm]{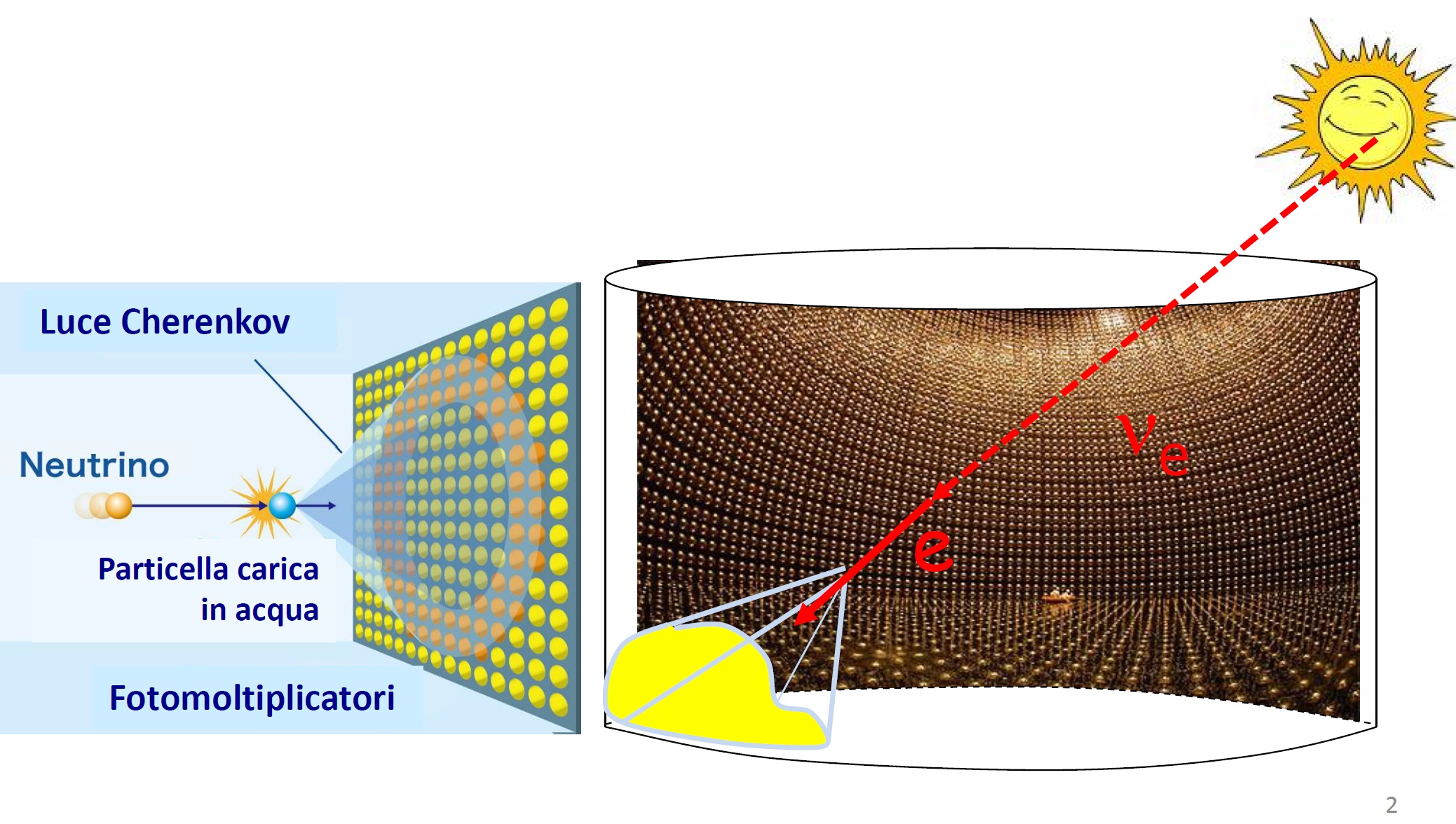}
\end{center}
\caption{\small Schema della rivelazione di neutrini solari da parte di Super-Kamiokande. Il neutrino proveniente dal Sole urta elasticamente un elettrone nel rivelatore. Il processo del $^8$B nel Sole produce $\nu_e$ (in viola nella Fig. \ref{fig:sole}) capaci di fornire più di 5 MeV di energia all'elettrone colpito, che si muove quindi in acqua con velocità maggiore della luce. L'acqua ha indice di rifrazione $n=1.33$ e la luce si propaga con velocità $v=c/n$.
Per questo motivo, l'elettrone induce radiazione luminosa per \textit{effetto Cherenkov}. Questa luce viene rivelata dai PMT, permettendo la determinazione dell'energia e della direzione dell'elettrone.  }
\label{fig:nu-sun}
\end{figure}

Il fatto che vi sia correlazione tra la direzione del Sole e quella di emissione dell'elettrone è fondamentale (Fig. \ref{fig:nu-sun}). Si noti infatti che il processo (\ref{eq:ES}) è indistinguibile dal fondo dovuto al decadimento di un nucleo radioattivo presente nell'acqua. 
Ma gli eventi di fondo sono diretti in modo casuale, mentre il segnale cercato produce un eccesso in corrispondenza della direzione del Sole.
Il numero di eventi contati in eccesso rispetto al fondo e imputabili all'arrivo di neutrini dal Sole corrisponde solo al $\sim$50\% degli eventi aspettati.

La combinazione dei risultati degli esperimenti sinora citati indica che ``mancano neutrini'' provenienti dal Sole rispetto ai modelli teorici. Tuttavia, nessuno di questi esperimenti ha potuto provare in maniera conclusiva che la mancanza di $\nu_e$ solari era dovuta al fenomeno delle oscillazioni. 

L'esperimento SNO (Sudbury Neutrino Observatory), che ha preso dati dal 1999 sino al 2006 in Canada, ha finalmente fornito questa prova in maniera decisiva. Per questo motivo, insieme al già menzionato T. Kajita, il premio Nobel 2015 è stato attribuito al suo portavoce, Arthur B. McDonald.
SNO era un esperimento capace di rivelare la luce Cherenkov emessa dalle particelle cariche propagantesi nel rivelatore, riempito con \( 1000\textrm{ t} \) di acqua pesante (\( D \)\( _{2} \)O) e circondato da 1500 t di acqua normale che fungeva da schermo. 
Il fatto che SNO usasse acqua pesante permetteva di rivelare non solo neutrini tramite il processo (\ref{eq:ES}) ma anche tramite:
\begin{equation} \label{eq:CC}
CC:\quad\quad \nu _{e}+n\rightarrow e^-+p \ .
\end{equation} 
Questo processo di interazione a corrente carica (CC) del $\nu_e$ sul neutrone è il corrispondente alla reazione (\ref{eq:8-nu1}) del $\overline \nu_e$ su $p$. 
Mentre per quest'ultimo processo protoni liberi sono disponibili (per esempio, l'idrogeno in H$_2$O), il processo (\ref{eq:CC}) \`{e} possibile solo su neutroni intrappolati nei nuclei, poiché neutroni liberi non sono disponibili in natura (o almeno, non sono disponibili in grande numero e per lungo tempo, visto che la loro vita media è di circa 15 minuti).
Il deutone (il nucleo di deuterio formato dal legame di un $p$ e $n$) contiene il neutrone su cui può avvenire la reazione (\ref{eq:CC}).
Con l'aiuto della tabella periodica degli elementi, potete facilmente convincervi che le reazioni (\ref{eq:Cl}) e (\ref{eq:Ga}) corrispondono esattamente alla (\ref{eq:CC}) con uno dei neutroni del nucleo del cloro e gallio, rispettivamente.

In aggiunta, nel deuterio presente nell'acqua pesante, pu\`{o} avvenire per tutti i tipi di neutrini la reazione in cui il deutone $d$ viene dissociato nei suoi costituenti:
\begin{equation}\label{eq:NC}
NC:\; \; \; \; \; \; \; \; \nu _{x}+d\rightarrow \nu _{x}+p+n,\; \; \; \; \; \nu _{x}=\nu _{e},\nu _{\mu },\nu _{\tau } \ .
\end{equation}
A seguito della dissociazione del $d$ in $p$ e $n$, un fotone di energia $\sim 2$ MeV \`{e} emesso. Nel rivelatore SNO era disciolto un sale che aumenta la probabilit\`{a} di cattura del neutrone, esattamente come nell'esperimento di Cowans e Raines, \S \ref{sec:sto1}. 
Il \( \gamma  \) da 8 MeV emesso dopo la cattura neutronica d\`{a} luogo a una coppia di $ e^{+}e^{-}$, oppure cede parte della sua energia ad un elettrone atomico via effetto Compton. Le particelle cariche producono luce Cherenkov e possono essere rivelate.

I modelli teorici basati sul Modello Solare Standard \cite{bahcall} prevedono un flusso di $\nu_e$ \textit{partenti dal Sole} con energia capace di rilasciare un segnale in SNO pari a:
\begin{equation}
\phi_{tot}(\nu)^{SSM} = (5.6\pm 0.8) \times 10^6\ \textrm{cm}^{-2} \textrm{s}^{-1}
\label{Eq:sno_2}
\end{equation}
Tramite la reazione (\ref{eq:NC}) si pu\`{o} misurare il flusso incidente totale dei neutrini solari, \( \nu _{e}+\nu _{\mu }+\nu _{\tau } \). Tramite la reazione (\ref{eq:CC}), si pu\`{o} invece misurare solo il flusso dei \( \nu _{e} \). 
Il numero totale di neutrini dal Sole \textit{misurati sulla Terra} da SNO tramite la reazione (\ref{eq:NC}) \`{e}:
\begin{equation}
\phi(\nu_e+\nu_\mu+\nu_\tau)^{SNO} = (5.25\pm 0.16_{stat} \pm 0.13_{sist}) \times10^6\ \textrm{cm}^{-2} \textrm{s}^{-1} \ .
\label{Eq:sno_3}
\end{equation}
SNO ha inoltre misurato il rapporto tra il flusso di neutrini elettronici e il flusso totale di neutrini, che risulta:
\begin{equation} 
R={\phi(\nu_e)\over \phi (\nu _{e} + \nu_\mu + \nu_\tau) } = 0.32 \pm  0.03 \ .
\label{Eq:sno_1}
\end{equation}
Questo risultato indica chiaramente che $\phi(\nu_\mu+\nu_\tau)$ \`{e} non nullo ed anzi corrisponde ai 2/3 di tutti i neutrini prodotti dal Sole,  e fornisce una prova definitiva del fatto che una parte dei neutrini elettronici solari, nel loro tragitto verso la Terra, cambia sapore. 
Il numero \textit{totale} di neutrini solari misurato (\ref{Eq:sno_3}) è quindi entro gli errori compatibile con quello (\ref{Eq:sno_2}) previsto dalla teoria.

Nel 2007 ai Laboratori del Gran Sasso \`{e} entrato in funzione Borexino, che ha iniziato a misurare i neutrini dovuti ai diversi processi mostrati in Fig. \ref{fig:sole}. Borexino usa scintillatore li\-qui\-do e la rivelazione dei neutrini avviene tramite l'urto elastico sull'elettrone (\ref{eq:ES}). 
Borexino ha raggiunto un cos\`{i} elevato grado di purificazione dalla radioattivit\`{a} da poter misurare il flusso di neutrini dovuti ai diversi processi mostrati nella figura (\ref{fig:sole}b). In particolare, Borexino ha misurato neutrini monocromatici ($E_\nu=0.862$ MeV) provenienti dalla cattura elettronica del $^7Be$, e quelli provenienti dal processi $pp$, $^8B$, e $pep$.

I risultati di SNO, di Borexino e degli altri esperimenti citati indicano che: 
\begin{enumerate}
\item le reazioni che avvengono all'interno del Sole e che permettono di mantenere le stelle in vita per miliardi di anni sono ben compresi; 
\item il flusso di neutrini elettronici \`{e} significativamente ridotto (tra 1/2 e  1/3) rispetto quanto aspettato;
\item il flusso totale di neutrini corrisponde a quanto previsto dal modello solare, in quanto i $\nu_e$ mancanti si sono trasformati durante il viaggio in neutrini di altri sapori ($\nu_\mu, \nu_\tau$). Questi ultimi non lasciano un segnale se non si utilizza una tecnica sperimentale analoga a quella usata da SNO.
\item quando si tiene conto delle \textit{oscillazioni dei neutrini}, il numero di $\nu_e$ misurati e attesi coincidono entro gli errori in ogni esperimento considerato.
\end{enumerate}
Quindi, mentre da un lato la misura del flusso di neutrini solari conferma le nostre conoscenze dell'astrofisica stellare, dall'altro ci ha permesso di capire una cosa fondamentale sui neutrini, ossia che si tratta di particelle di massa non nulla (\S \ref{sec:osci}).

\section{``Morte''. Ossia, i neutrini e la fine di una stella\label{sec:morte}}
La nucleosintesi nelle stelle (ossia la fusione di nuclei leggeri per ottenere nuclei più pesanti) è esotermica (rilascia energia) sino alla produzione del ferro. 
Quando all'interno di una stella cessa la produzione di energia dovuta alla fusione, si arresta la produzione dei fotoni che mantengono la stella in equilibrio contro la forza di gravità: la stella subisce un \textit{collasso gravitazionale}. 

\begin{figure}[tb]
\begin{center}
\includegraphics[width=9.5cm]{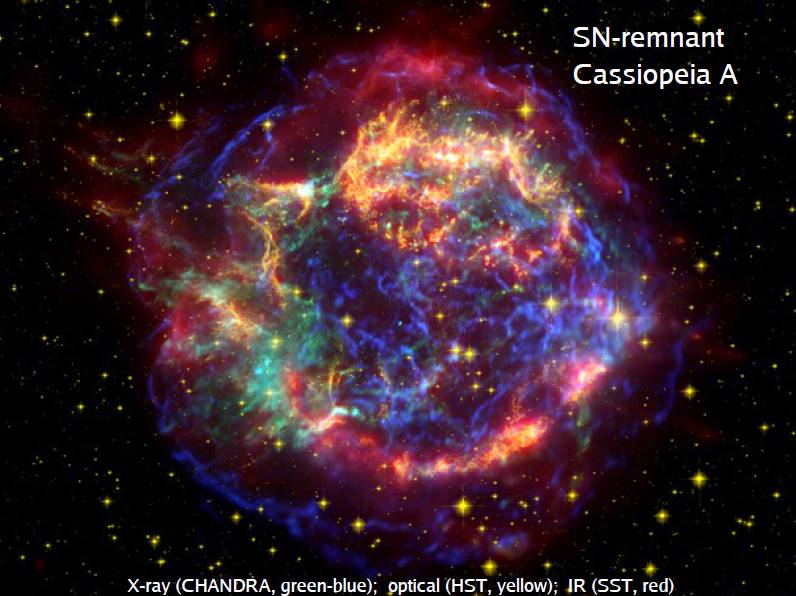}
\end{center}
\caption{\small Immagine del resto della SN1572 esplosa nella costellazione di Cassiopeia ed osservata da Tycho Brahe. L'immagine è formata dalle osservazioni effettuate nel visibile dal telescopio spaziale Hubble (giallo), nei raggi X dal satellite Chandra (blu e verde) e nell'infrarosso dallo Spitzer Space Telescope (rosso). Il materiale che rilascia l'immagine corrisponde ad un'onda di shock che si allontana con velocità $v/c\sim 10^{-2}$ dal centro di quella che era la stella. La velocità si evince anche dalle dimensioni della nube (qualche pc) e dai $\sim$ 500 anni trascorsi dall'esplosione. Crediti: X-ray: NASA/CXC/SAO; Ottico: NASA/STScI; Infrarosso: NASA/JPL-Caltech}
\label{fig:sn1572}
\end{figure}

Il collasso gravitazionale stellare è uno dei processi che dà luogo alle cosiddette esplosioni di \textit{supernovae} (SNe). Si tratta di un fenomeno spettacolare e rarissimo da vedere a occhio nudo, ma abbastanza frequente quando si utilizzano potenti telescopi. Quando avviene una SN visibile nella nostra Galassia\footnote{detta anche \textit{Via Lattea}.}, appare una sorgente luminosa che per alcuni giorni o settimane supera quella di ogni altra stella. 
Le ultime SN osservate ad occhio nudo si sono avute l'11 novembre 1572 (riportata dallo scienziato danese Tycho Brahe), e nell'ottobre del 1604 (SN detta anche di Keplero, che sulle osservazioni effettuate scrisse un libro). La Fig. \ref{fig:sn1572} mostra come appare la SN di Tycho ai nostri giorni.
L'avvento degli strumenti moderni e soprattutto della fotografia ha permesso, dalla fine del 1800, di osservare SNe in altre galassie. In questi casi, la SN supera talvolta la luminosità dell'intera galassia che ospita la stella morente.
Basandosi su queste osservazioni, gli astronomi stimano che dovrebbero avvenire al più tre esplosioni di supernova per secolo nella Via Lattea. 
Il fatto che l'occhio umano ne abbia viste molte meno è dovuto all'assorbimento della luce da parte del materiale interstellare. Questo fenomeno non fermerebbe i neutrini, la cui rivelazione dunque annuncerebbe la morte di una stella di grande massa nella Via Lattea.

Come detto in \S \ref{sec:vita}, il Sole nei prossimi 5 miliardi di anni fonderà nel suo centro nuclei H in He e successivamente He in C.
Stelle più grandi del Sole, in virtù della loro maggiore pressione gravitazionale, sono in grado di attuare dei cicli di fusione che dal C portano alla produzione di elementi sempre più pesanti. 
Esse evolvono dunque in una struttura ``a cipolla'', con una sequenza di
strati concentrici dentro i quali avvengono reazioni nucleari di natura diversa. Lo strato più esterno è costituito da idrogeno; procedendo verso il centro della stella, si trovano in sequenza gli strati di elio, carbonio, ossigeno, e con il nucleo centrale composto ferro.
Solo le stelle con massa $M>8 M_\odot$ riescono ad avere una pressione nel core così elevata da sintetizzare ferro.
Nuclei più pesanti del ferro non possono essere sintetizzati senza consumare energia. Quindi, quando all'interno di una stella si sintetizza ferro, cessa la produzione di energia e la stella subisce il  {collasso gravitazionale}: davvero \textit{il più grande spettacolo dopo il Big Bang}.


Cosa c'entrano i neutrini in tutto questo? Di nuovo, entra in gioco la nostra necessità di comprendere come avvengono i fenomeni in natura e il principio di conservazione dell'energia.
Infatti, le misure di \textit{ogni forma di energia osservabile} dalla SN indicano una carenza di circa il 99\% dell'energia a disposizione.
Vediamo qualche dettaglio del modello, e come è stato risolto il nuovo mistero dell'energia mancante.
 
Quando nel nucleo della stella si è formato il ferro, gli stati esterni continuano a ``precipitare'' verso il centro. A questo punto, il nucleo stellare compresso diventa così denso che i protoni nel ferro iniziano a catturare gli elettroni circostanti\footnote{La materia in queste situazioni estreme è completamente ionizzata. Tuttavia, la materia è complessivamente neutra, e quindi se in una certa regione di spazio ci sono nuclei, ci sono anche elettroni.} e si trasformano in neutroni tramite la reazione:
\begin{equation} \label{eq:cattura}
e^- + p\rightarrow \nu _{e} +n 
\end{equation} 
(si noti che essa è l'inversa della Eq. \ref{eq:CC}). Inizia così la formazione di quella che diventerà una \textit{stella di neutroni}.
La stella di neutroni è uno stato \textit{degenere} della materia. Degenere significa che può verificarsi solo in determinate condizioni, che sono però studiabili con la fisica quantistica. Questo stato degenere arriva a una densità di circa $3\ 10^{14}$ g/cm$^3$. Un centimetro cubo di una stella di neutroni ha una massa di un terzo di miliardo di tonnellate!
Il fisico indiano S. Chandrasekhar (Nobel nel 1983)\footnote{Il satellite della NASA per lo studio dei raggi X, emessi anche dalle stelle di neutroni ruotanti velocemente su se stesse (\textit{pulsar}), è stato chiamato Chandra proprio in onore del fisico indiano.} riuscì a determinare a metà del secolo scorso che una stella di neutroni deve avere massa $M_{NS}$ pari a circa 1.5 $M_\odot$ in un raggio $R_{NS}$ dell'ordine di $\sim$ 10 km.

La meccanica newtoniana permette di determinare l'energia di legame di un oggetto legato dalla gravità. Il calcolo si basa sul lavoro compiuto dalla forza di gravità per formare un oggetto di massa $M$ e raggio $R$ ed è semplice per uno studente contemporaneo\footnote{Newton contribuì ad inventare il calcolo integrale anche per poter risolve questo problema.}.  Questa energia potenziale è pari a 
$ |U_G|= {3 G M^2 \over 5 R} $, dove $G$ è la costante di gravitazione universale.
Quando si forma la stella di neutroni, la variazione di energia potenziale è pari a
\begin{equation}\label{eq:du}
\Delta U = |\frac{3GM^2}{5R} - \frac{3GM_{NS}^2}{5R_{NS}} | \simeq \frac{3GM_{NS}^2}{5R_{NS}} \simeq 3\times 10^{46} \textrm{ J} \ ,  
\end{equation}
relazione praticamente indipendente dal valore iniziale di $M$, perché $R\gg R_{NS}$.

Quando la stella si disintegra sotto forma di supernova, l'energia rilasciata deve corrispondere al valore sopra riportato.
Le osservazioni mostrano però che la radiazione emessa contribuisce solo per $10^{-4}$ di $ \Delta U$. All'energia cinetica di tutto il materiale emesso, che si propaga nella galassia con velocità media circa 1/100 delle velocità della luce (si veda anche la Fig. \ref{fig:sn1572}), compete solo l'1\% di $\Delta U$. 
Per molto tempo questa sbilancio energetico tra teoria e osservazione nei collassi gravitazionali fu un affascinante enigma astrofisico.
Finalmente, a partire dalla fine degli anni '60 si iniziò a ipotizzare che l'energia mancante finisse sotto forma di neutrini, e che questi avessero un ruolo fondamentale nel collasso gravitazionale stellare.
A partire dagli anni '80, cominciarono a svilupparsi anche simulazioni per cercare di riprodurre il fenomeno al calcolatore.

Anche tramite queste simulazioni, sappiamo che una parte dell'energia di legame gravitazionale ($\sim $10\%) viene persa nel meccanismo di formazione della stella di neutroni, tramite l'emissione dei neutrini nella (\ref{eq:cattura}).
Questo processo (chiamato di \textit{neutronizzazione}) avviene in un tempo brevissimo, circa 1/100 di secondo. 
Ora nella zona centrale di quella che era la nostra stella vi è un oggetto così denso che la caduta del materiale comincia ad arrestarsi.
Arrestandosi, la densità di materia subito all'esterno del core comincia a diventare così elevata che diventa persino difficile la fuga dei neutrini, che continuano a formarsi attraverso diversi processi fisici. 
La densità dei neutrini comincia ad essere così elevata che (come quando si gonfia troppo un palloncino) il sistema esplode. La materia viene espulsa verso l'esterno (si ha appunto l'esplosione della supernova) e i neutrini sono liberi di vagare nell'universo. Di nuovo, tutto il processo dura non più di qualche decina di secondi.
Le stime teoriche mostrano che l'energia che i neutrini trasportano via compensano il rimanente dell'energia mancante stimato dalla (\ref{eq:du}).

Questo modello teorico sembrava quasi bizzarro agli occhi degli astrofisici: insignificanti neutrini che si portano via la quasi totalità dell'energia di legame di una stella, che assurdità! E non solo: il modello prevede che neutrini e antineutrini di ogni sapore e in egual numero si formino nel processo.
Come nel caso dell'ipotesi di Pauli, la conferma sperimentale di questo modello sembrava irraggiungibile\footnote{Vi lascio il seguente esercizio: supponete di avere una esplosione di SN di 10 $M_\odot$ nel centro della nostra Galassia, a 8.5 kpc dalla Terra. Sapendo che 1 pc=$3\ 10^{18}$ cm, e che i neutrini emessi hanno energia di circa $10^{-13}$ J, determinare il flusso di totale di neutrini sulla Terra. Vi fornisco la risposta: circa $10^{11}$ per cm$^{2}$, durante un intervallo di circa 30 s. Confrontate questo flusso con quello proveniente dal reattore nell'esperimento di Cowan e Reines, \S \ref{sec:sto1}.}.

\subsection{Esperimenti per la rivelazione di neutrini dalle SN \label{sec:morteexp}}

Negli anni '80, come abbiamo detto in \S \ref{sec:under}, erano nati i primi esperimenti underground per la ricerca del decadimento del protone. 
Poiché il decadimento del protone richiedeva apparati sperimentali della scala di $>1$ kton, questi erano molto più grandi dell'esperimento di Cowan e Reines.
Alcuni di questi esperimenti erano estremamente versatili, e sarebbero stati in grado di rivelare le interazione degli $\overline \nu_e$ in arrivo da una SN, se mai ce ne fosse stata una nella Via Lattea. Il problema, come visto, è che il numero di SN previste nella nostra Galassia è, al più, di 2-3 per secolo!
I ricercatori non avevano quindi molte speranze. Tuttavia, il 23 Febbraio 1987 apparve agli occhi di un astrofilo una supernova nella Grande Nube di Magellano. Questa è una piccola galassia satellite della Via Lattea, ad una distanza da noi di circa 50 kpc.
I responsabili dei rivelatori dei neutrini furono subito allertati, e si iniziò quasi immediatamente a cercare nei dati memorizzati il possibile segnale dovuto alla SN. 
Successivamente, fu anche possibile rintracciare negli archivi fotografici la posizione dell'esplosione e localizzare la stella progenitrice, una supergigante blu di massa 20 $M_\odot$. 

\begin{figure}[tb]
\begin{center}
\includegraphics[width=9.5cm]{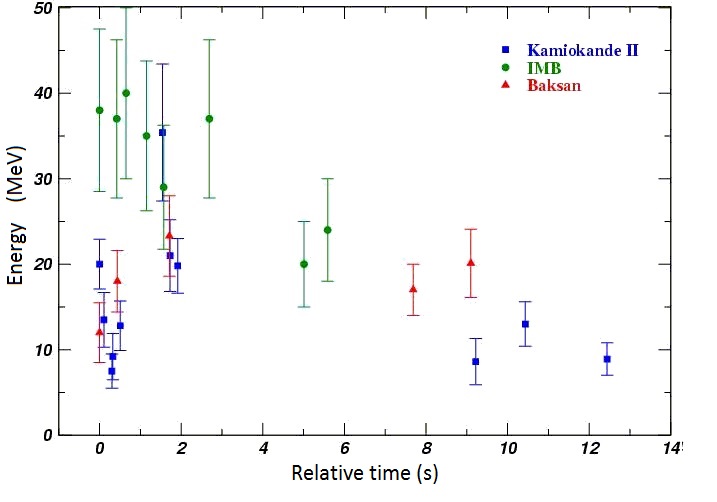}
\end{center}
\caption{\small Tempi (relativi) di arrivo ed energia degli eventi dovuti all'interazione di $\overline \nu_e$ in arrivo dalla SN1987A e osservati dagli esperimenti Kamiokande, IMB e Baksan. L'istante del primo evento per i tre esperimenti è stato arbitrariamente posto a t=0.}
\label{fig:sn1987}
\end{figure}

I due esperimenti attivi più grandi erano capaci di rivelare la luce Cherenkov nell'acqua prodotta dal decadimento beta inverso (\ref{eq:8-nu1}), ossia la stessa reazione utilizzata per la scoperta del neutrino. 
Kamiokande (in Giappone) era un cilindro con un volume fiduciale di circa 2 kton di acqua (grossomodo, quanta quella contenuta in una piscina olimpica) mentre l'esperimento IMB (nell'Ohio, USA) era un enorme parallelepipedo che utilizzava circa 5 kton di acqua.
Gli esperimenti osservavano la luce emessa dai positroni nella reazione (\ref{eq:8-nu1}) per mezzo dei fotomoltiplicatori posti sulle pareti. Quello in Giappone aveva PMT più grandi e sensibili.
I due esperimenti rivelarono 11 (Kamiokande) e 8 (IMB) interazioni di antineutrini, in un intervallo temporale di 10 s. Un terzo esperimento in Russia (Baksan), più piccolo, utilizzava una tecnica diversa per misurare sempre la reazione (\ref{eq:8-nu1}) e rivelò anch'esso circa 5 eventi (qualche evento poteva essere dovuto a fondo spurio). La distribuzione delle energie dei positroni misurati dai tre esperimenti e il relativo tempo di arrivo sono riportati in Fig.  \ref{fig:sn1987}.

Il segnale dei neutrini misurato dai tre esperimenti (il numero di eventi, le energie e tempi di arrivo come riportati nella figura) in occasione della SN 1987A ha confermato il modello di una esplosione di una stella di molte volte la massa solare.
Il raffreddamento attraverso l'emissione dei neutrini ha trasportato via gran parte dell'energia di legame gravitazionale della stella esplosa nella Grande Nube di Magellano. 

I neutrini da supernova lasciano la regione di produzione poco prima della luce; la loro rivelazione può quindi dare in anticipo un allarme agli astronomi che una supernova sta per apparire. Ad esempio Super-Kamiokande (oggi ancora in funzione) dovrebbe registrare diverse migliaia di eventi da una supernova nel centro della Via Lattea, rivelazione che permetterebbe di localizzare la SN nel cielo con una risoluzione di qualche grado.
Un apposito sistema di allarme connette gli esperimenti per neutrini a quelli ottici per dare un precoce avviso agli astronomi dell'evento di una supernova nella nostra galassia.
Ulteriori dettagli nel Cap. 12 di \cite{spurio} e in \cite{mirizzi,schol}.


\section{``Miracoli''. Ossia, i neutrini e cosa succede dopo la morte di una stella\label{sec:miracoli}}

L'osservazione dei neutrini solari, e di quelli emessi nell'esplosione della supernova 1987A, ha permesso di rispondere ad alcuni interrogativi riguardanti i processi che hanno luogo all'interno di oggetti stellari e hanno acceso l'interesse verso i neutrini di alta energia emessi da sorgenti cosmiche.

Come abbiamo introdotto in \S \ref{sec:under}, la Terra è continuamente bombardata da raggi cosmici (RC), accelerati da oggetti astrofisici \cite{deangelis}. Sperimentalmente sono stati osservati RC sino ad energie di $ 10^{20}$ eV; si tratta di una energia circa 10 milioni di volte più grande di quella raggiunta dal più potente acceleratore di particelle mai costruito dall'uomo, LHC al CERN.
Il Sole accelera protoni e altri nuclei fino ad energie dell'ordine di una decina di GeV (ossia, con energia cinetica pari circa 10 volte quella della massa a riposo del protone). 
Dinamicamente, non è possibile per le stelle ``in vita'' accelerare RC ad energie molto più grandi. I modelli indicano che solo oggetti prodotti ``dopo la morte'' di una stella possono presentare meccanismi di accelerazione necessari per spiegare il flusso di RC osservati.

Si pensa infatti che i resti di SN (come quelli mostrati in Fig. \ref{fig:sn1572}) possano accelerare protoni sino ad energie di un milione di GeV (=$ 10^{15}$ eV). 
Altri oggetti particolari nella nostra Galassia (pulsar giovani, sistemi binari di una stella e un buco nero\footnote{un buco nero può venire prodotto al posto di una stella di neutroni nel caso di un collasso gravitazionale di una stella molto massiva, ad esempio $M>20 M_\odot$.}) potrebbero accelerare RC sino a $ 10^{18\div19}$ eV. 
Oggetti extragalattici, quali i nuclei di galassie attivi (AGN) potrebbero essere le sorgenti dei RC di energia estrema. Un altro candidato extra-galattico è rappresentato dai getti di raggi gamma (gamma-ray burst, GRB), testimonianza diretta dell'esplosione di una stella gigantesca o della coalescenza di due oggetti compatti (ad esempio, della caduta di due stelle di neutroni, dopo aver spiraleggiato una sull'altra).

Tuttavia, a parte i RC dal Sole, le prove sperimentali del quadro sinora esposto sono lacunose. Il problema è che la Galassia (e anche le regioni di spazio tra le galassie) hanno dei campi magnetici che alterano la direzione di provenienza dei RC.
Quindi, osservare la direzione di provenienza di un RC non permette di individuare la sorgente che lo ha prodotto. Per questo, occorre utilizzare sonde neutre per verificare questi modelli. Le uniche sonde neutre provviste dalla natura capaci di giungere da regioni molto lontane dell'universo sono le particelle neutre stabili, ossia i fotoni e i neutrini\footnote{A questi si stanno aggiungendo le \textit{onde gravitazionali}. Ma questo è un altro capitolo dello studio dell'universo.}. Fotoni e neutrini sono prodotti a seguito dell'interazione dei RC con la materia (o il campo di radiazione) in prossimità della sorgente.

Esperimenti che misurano con precisione fotoni di altissima energia sono entrati in funzione nell'ultima decade. Fotoni di energia sino a $\sim 300 $ GeV sono rivelati con l'esperimento su satellite Fermi-LAT\footnote{Nome assegnato dalla NASA al satellite per l'astronomia dei raggi gamma lanciato nel 2008}; fotoni dalle migliaia di GeV sino a $\sim 100$ TeV (1 TeV = 1000 GeV) sono invece rivelati con particolari telescopi a Terra (si veda  \cite{funk} e i capitoli 8 e 9 di \cite{spurio} per una rassegna più dettagliata). 
Uno dei problemi dell'astronomia con raggi gamma, tuttavia, è il fatto che i fotoni possono essere assorbiti dalla materia che si frappone tra la sorgente e la Terra. 
I neutrini, invece, soffrono molto meno della presenza di materiale assorbente.

Una delle caratteristiche generali dei meccanismi di accelerazione astrofisici a cui sono sottoposte particelle cariche e stabili (ossia: protoni, elettroni e nuclei) è che il numero di particelle decresce al crescere della loro energia. 
Questo è vero non solo per le particelle accelerate, ma anche per quelle prodotte dalla loro interazione, quali i raggi gamma e i neutrini. 
Meccanismi di accelerazione sono stati ideati dall'uomo e utilizzati in laboratori e vari dispositivi terrestri: i tubo catodici (come quelli presenti nei vecchi apparati televisivi) accelerano elettroni sottoposti ad una differenza di potenziale. 
Un \textit{collider} (ossia, un acceleratore di particelle circolare) permette a particelle cariche di raggiungere energie elevatissime.
L'aumento di energia si ottiene facendole passare ciclicamente attraverso opportune differenze di potenziale. 
L'orbita circolare è mantenuta grazie alla presenza di campi magnetici.
Nel collider in ogni istante le particelle presenti si muovono in fase e con la stessa energia. 

Un ipotetico acceleratore astrofisico funziona per certi aspetti in modo differente. Il modello di accelerazione più accreditato (sviluppato per primo da E. Fermi \cite{fermiCR}. Ancora lui!) prevede che particelle possano essere accelerate da urti ripetuti con un'onda di shock (come quella prodotta da una supernova). 
Nell'urto, la particella accelera, come una pallina da tennis viene accelerata da un colpo di racchetta. La presenza di campi magnetici nella regione di accelerazione fa curvare la particella carica, che ritorna sull'onda di shock e, urtando di nuovo, aumenta ulteriormente la sua energia. 
Più elevato è il numero di urti ripetuti, maggiore è l'energia che la particella guadagna. Tuttavia, poiché i campi magnetici astrofisici sono molto più irregolari che quelli presenti in un collider terrestre, c'è una probabilità finita che, dopo ogni colpo, la particella esca definitivamente dalla regione in cui vi è l'onda di shock. Quindi il numero di particelle che subisce N urti consecutivi decresce al crescere di N. Il modello matematico di Fermi prevede che il numero di particelle con una certa energia decresca come
\begin{equation}
\label{eq:dnde}
{dN\over dE} \propto \biggl( {E\over E_0}\biggr)^{-2}
\end{equation}
ossia, se $10^6$ particelle raggiungono energia $E_0$, solo $10^4$ arrivano a energie $10\cdot E_0$,  100 a energie   $100\cdot E_0$, e una sola a energia $1000\cdot E_0$.

La stessa cosa vale per le particelle secondarie (raggi gamma e neutrini): poiché i neutrini sono prodotti dal decadimento di particelle cariche instabili (a loro volta prodotte dall'interazione di protoni, elettroni e nuclei accelerati dalle onde di shock), anch'essi hanno una identica dipendenza dall'energia, ossia $\propto E^{-2}$.

Dobbiamo quindi aspettarci neutrini di origine astrofisica non solo dovuti alla ``vita'' della stella (come i neutrini dell'ordine del MeV rivelati dal Sole) o testimoni dell'atto di ``morte'' della stella (come  i neutrini dell'ordine delle decine di MeV rivelati dalla SN1987A), ma anche neutrini di energia elevatissima prodotti ``miracolosamente'' dopo la morte delle stelle da stati degeneri della materia (buchi neri, stelle di neutroni,...). Tuttavia, il loro numero decresce molto al crescere di $E$ e occorrono quindi esperimenti di scala gigantesca per poterli rivelare.

\subsection{Esperimenti per la rivelazione di neutrini di energia elevatissima \label{sec:neutel}}

L'idea per la realizzazione di quelli che oggi chiamiamo \textit{telescopi di neutrini} fu del russo M.A. Markov \cite{markov}, che all'inizio degli anni '60 propose di porre un numero molto elevato di rivelatori ottici -fotomoltiplicatori (PMT) analoghi a quelli usati da SK, SNO o Borexino- sotto un grande spessore di acqua marina o di un lago, attrezzando un volume dell'ordine di 1 km$^3$. L'acqua avrebbe fornito il mezzo (gratuito) in cui i neutrini di altissima energia avrebbero potuto interagire; inoltre poiché l'acqua è trasparente, la luce emessa per effetto Cherenkov dalle particelle cariche prodotte dall'interazione sarebbe stata raccolta dai PMT. Infine, ponendo la strumentazione in profondità, si avrebbe avuto la schermatura necessaria per ridurre di molti ordini di grandezza la radiazione di luce solare e il flusso dei RC secondari.

La sfida offerta da questa proposta dal punto sperimentale è enorme. Occorre disporre di un numero di PMT pari a quelli di Super-Kamiokande, ma lungo stringhe o torri di 1 km di lunghezza e da immergere in profondità. 
Per avere una idea, se si vuole ``rendere attivo'' 1 km$^3$ di acqua disponendo una griglia di 10000 PMT egualmente distanziati, la loro distanza deve essere di circa 50 m uno dall'altro. 

Un primo tentativo per far funzionare una stringa sott'acqua iniziò negli anni '80 con una collaborazione russo-americana (DUMAND), che cercò di realizzare un esperimento a 4.5 km di profondità nel Pacifico, al largo delle isole Hawaii.
La tecnologia sottomarina dell'epoca non era sufficientemente avanzata e il tentativo fallì. Inoltre, le tensioni politiche tra sovietici e americani dopo l'invasione dell'Afghanistan da parte dei primi fecero sì che l'amministrazione Reagan vietasse collaborazioni scientifiche tra americani e russi. 
I primi iniziarono così a pensare di costruire un telescopio di neutrini sotto il ghiaccio dell'Antartide (lo spessore di ghiaccio è di circa 2.5 km), mentre i russi iniziarono gli studi per la realizzazione di un telescopio nel lago Baikal, alla profondità di 1.1 km.
Per loro conto, gli europei avevano intanto iniziato le attività di ricerca e sviluppo per costruire un telecopio di neutrini nel mar Mediterraneo.

\begin{figure}[tb]
\begin{center}
\includegraphics[width=10.5cm]{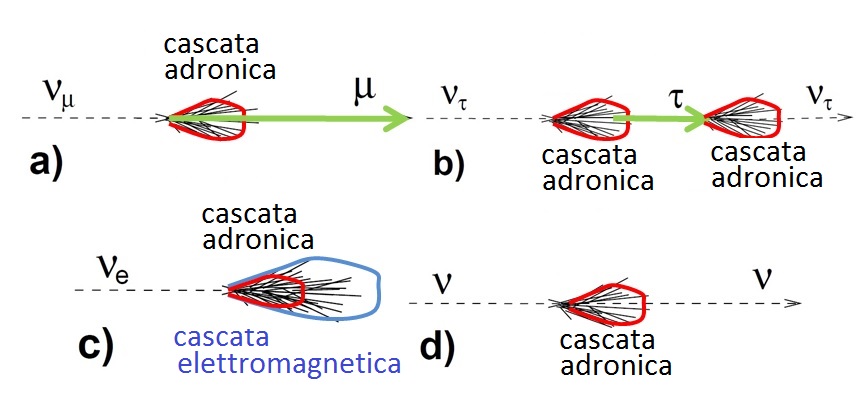}
\end{center}
\caption{\small Interazioni dei neutrini di diversi sapori: 
(a) interazione a corrente carica di un $\nu_\mu$ con produzione di un muone; 
(b) interazione a corrente carica di un $\nu_\tau$ con produzione di un tau; 
(c) interazione a corrente carica di un $\nu_e $ con produzione di un elettrone che induce una cascata di particelle;
(d) interazione a corrente neutra di neutrino di ogni sapore.}
\label{fig:topo}
\end{figure}
Ci si aspetta che le sorgenti astrofisiche, tenendo conto della loro distanza e delle oscillazioni dei neutrini (vedere \S \ref{sec:osci}), inviino $\nu_e, \nu_\mu, \nu_\tau$ (e i rispettivi antineutrini) in egual numero.
Alle altissime energie dei neutrini e antineutrini in oggetto, essi possono interagire indifferentemente su protoni e neutroni ($N$) del mezzo. 
Quando interagiscono per mezzo delle cosiddette \textit{interazioni a corrente carica}, nello stato finale viene prodotto il leptone carico corrispondente al neutrino interagente:
\begin{equation}\label{eq:nuCC}
\nu_\ell + N  \rightarrow \ell + X \quad\quad \ell=e,\mu,\tau 
\end{equation}
il leptone $\ell$ trasporta in media oltre il 50\% dell'energia del neutrino incidente, mentre la rimanente energia è usata per creare molte particelle \textit{adroniche}, ossia composte da quark e indicate con $X$ in (\ref{eq:nuCC}).
Nella regione interna di questi giganteschi esperimenti non è ovviamente possibile creare un campo magnetico che separi leptoni negativi ($\ell$)  da quelli positivi ($\overline \ell$), e non si possono quindi distinguere interazioni di neutrini da quelle di antineutrini.

I tre leptoni carichi prodotti dall'interazione di neutrino producono segnali differenti in un telescopio. L'elettrone prodotto dal $\nu_e$ è un particella leggera ed estremamente energetica che immediatamente produce raggi $\gamma$ per radiazione di frenamento; a loro volta i $\gamma$ creano coppie $e^+e^-$. In breve, si viene a creare una cascata di un numero elevatissimo di fotoni, elettroni e positroni, chiamata \textit{cascata elettromagnetica}, Fig.  \ref{fig:topo}c. 
Le particelle nella cascata crescono di numero, arrivano ad una numerosità massima e poi cominciano a decrescere. Tutto questo in una regione di spazio che si estende in acqua per una decina di metri.  Il numero di particelle nella cascata è proporzionale all'energia del primo leptone prodotto dal neutrino, e quindi proporzionale all'energia del neutrino incidente.

Le particelle cariche nella cascata elettromagnetica producono luce Cherenkov in gran parte del loro percorso, una frazione della quale viene raccolta dai PMT. 
La cascata può essere approssimata come un ellissoide dal diametro maggiore molto minore della distanza tra i sensori ottici. 
I segnali raccolti da tutti i PMT coinvolti nell'evento vengono usati per ricostruire l'energia del neutrino (proporzionale alla quantità di luce raccolta) e la direzione di provenienza del neutrino (coincidente col diametro maggiore dell'ellissoide). 
Questa topologia di eventi permette una stima piuttosto accurata dell'energia del neutrino; la direzione di provenienza è stimata più approssimativamente, essendo ricostruita con precisione di 3$^\circ$-4$^\circ$ nei casi migliori.
Per avere un'idea, le dimensioni angolari della luna nel cielo corrispondono a circa 0.5$^\circ$.

Nel caso di un muone prodotto dal $\nu_\mu$ la situazione è completamente differente. Il muone ha proprietà analoghe a quelle dell'elettrone, a parte una massa 200 volte maggiore. Per questo, il muone non emette radiazione di frenamento e può propagarsi praticamente lungo una retta per distanze anche dell'ordine di parecchi chilometri, Fig.  \ref{fig:topo}a. Durante la propagazione, il muone emette luce Cherenkov che viene raccolta da diversi PMT disposti in prossimità del suo passaggio. I segnali permettono di ricostruire con relativa precisione la direzione del muone (e quindi del neutrino incidente che lo ha prodotto), arrivando a 0.2$^\circ$-0.3$^\circ$. Viceversa, l'energia del neutrino viene stimata con precisione molto inferiore rispetto al caso della cascata elettromagnetica.

Il tau prodotto dal $\nu_\tau$ ha una situazione ibrida tra i due casi precedenti: in talune situazioni può essere visto principalmente tramite la cascata elettromagnetica; in talaltre, poiché il tau ha un comportamento analogo al muone, può essere visto come una traccia, Fig.  \ref{fig:topo}b. 

Infine, neutrini di tutti e tre i sapori possono interagire con protoni e neutroni dei nuclei in un processo detto a \textit{corrente neutra} 
\begin{equation}\label{eq:nuNC}
\nu_\ell + N  \rightarrow \nu_\ell + X \quad\quad \ell=e,\mu,\tau 
\end{equation}
in cui parte dell'elevata energia del neutrino è usata per creare molte particelle  {adroniche}.
La maggior parte di queste particelle sono instabili e decadono in altre particelle. Tutto questo produce una \textit{cascata adronica},  Fig.  \ref{fig:topo}d. Tipicamente, anche una cascata adronica si sviluppa ed esaurisce in acqua in una decina di metri. 
La luce prodotta dalle cascate adroniche (presenti non solo nel caso delle correnti neutre, ma anche prodotta dalle particelle $X$ nelle interazioni di corrente carica, Eq. \ref{eq:nuCC}) viene anch'essa rivelata, ma è in generale minore di quella prodotta dagli sciami elettromagnetici. 

Per l'astronomia di neutrini \cite{chiarusi,becker}, ossia la possibilità di identificare con precisione la posizione delle sorgenti, l'osservazione dei muoni rappresenta quindi il canale privilegiato.
In tutte le topologie, i soli eventi sicuramente indotti da neutrini sono quelli diretti verso l'alto. I muoni atmosferici diretti verso il basso rappresentano infatti una contaminazione che oscura completamente il segnale. Anche negli eventi verso l'alto, vi è il fondo irriducibile dei neutrini atmosferici. Questo fondo è completamente isotropo e decresce molto più velocemente del segnale al crescere dell'energia. 

I neutrini di origine astrofisica possono essere selezionati rispetto al fondo in due modalità: o evidenziando un eccesso di eventi da una data direzione (principalmente, usando gli eventi $\nu_\mu \rightarrow \mu \rightarrow $ traccia) o evidenziando un eccesso di eventi di più alta energia rispetto a quanto aspettato nei neutrini atmosferici (usando eventi $\nu_e \rightarrow e \rightarrow$ cascata elettromagnetica).
In talune situazioni, è possibile selezionare cascate indotte da eventi di alta energia anche diretti verso il basso, ma con risoluzione angolare piuttosto bassa.

Attualmente, i due principali telescopi di neutrini sono IceCube in Antartide e ANTARES nel mar Mediterraneo. 
\begin{figure}[tb]
\begin{center}
\includegraphics[width=9.5cm]{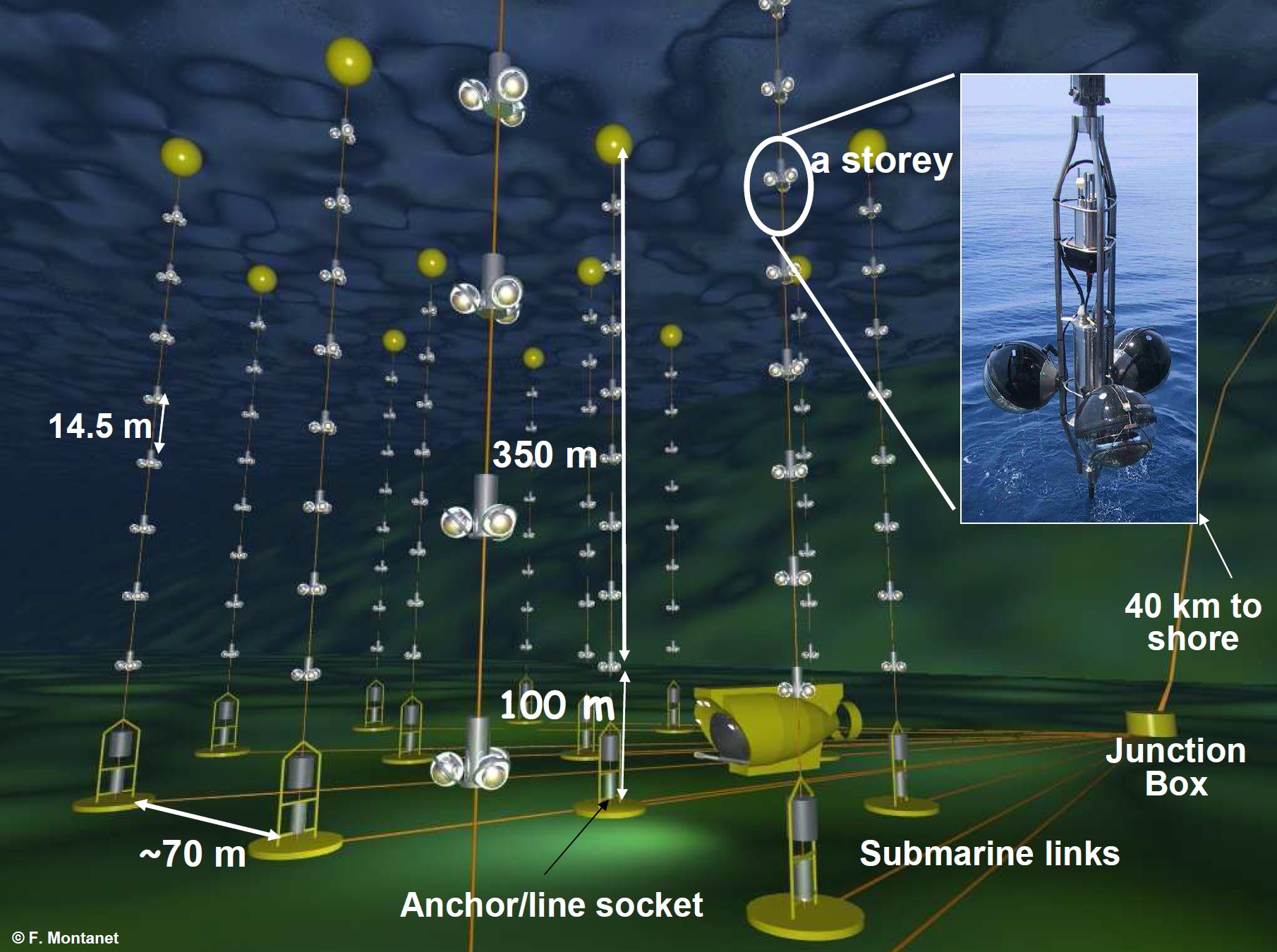}
\end{center}
\caption{\small Disegno del telescopio di neutrini sottomarino ANTARES. Esso consiste di 875 PMT, raggruppati in tripletti (\textit{storey}), fissati a 12 cavi elettro-ottici. Uno di questi \textit{storey} è mostrato nell'inserto.
 L'orientazione dei PMT è stata scelta per ottimizzare la rivelazione di neutrini provenienti dal basso.
I cavi, saldamente ancorati sul fondo del mare e tenuti in tensione da apposite boe in cima, svolgono il triplo ruolo di reggere meccanicamente i PMT, di fornire ai medesimi e all'elettronica associata la tensione elettrica necessaria e di trasmettere verso la stazione di controllo posta a terra i segnali registrati dai PMT. A tale scopo è adoperato 
un cavo sottomarino, lungo circa 40 km che si connette alla \textit{junction box}. }
\label{fig:antares}
\end{figure}

ANTARES \cite{antares}, costruito e mantenuto in funzione da un'ampia collaborazione di istituzioni e università soprattutto europee, è installato al largo di Tolone (Francia) a circa 40 km dalla costa, ad una profondità di 2400 metri (Fig. \ref{fig:antares}). 
L'installazione e le eventuali operazioni di riparazione del rivelatore richiedono l'utilizzo di navi e di sottomarini teleguidati.
I PMT sono racchiusi in sfere di vetro resistenti alla pressione (che arriva a $\sim$250 atmosfere alla profondità massima). 
Le correnti marine possono spostare lateralmente le boe e quindi tutti i PMT, anche di diversi metri. 
Le procedure per determinare il corretto posizionamento dei sensori istante per istante e quelle di calibrazione assumono per tale esperimento importanza fondamentale.
Infine, la radioattività nell'acqua di mare (dovuta al $^{40}$K) e  la presenza (anche a grandi profondità) di organismi marini che emettono piccoli impulsi luminosi rappresentano difficoltà addizionali per l'elettronica di selezione del segnale e il sistema di acquisizione dati.

ANTARES (costruito tra il 2006 e il 2008) ha dimostrato che si possono vincere tutte le difficoltà tecniche che si sono presentate nella realizzazione e nel funzionamento di un telescopio di neutrini nelle profondità marine. Dal 2008 ANTARES ha continuamente preso dati, studiano il cielo Sud con alta efficienza. Il suo spegnimento è previsto per la fine del 2017, quando un nuovo, più grande ed efficiente telescopio di neutrini (denominato KM3NeT \cite{km3}) entrerà in funzione al largo delle coste della Sicilia. Anche KM3NeT è un progetto largamente europeo, che si basa sull'esperienza accumulata da ANTARES.

\begin{figure}[tb]
\begin{center}
\includegraphics[width=12.0cm]{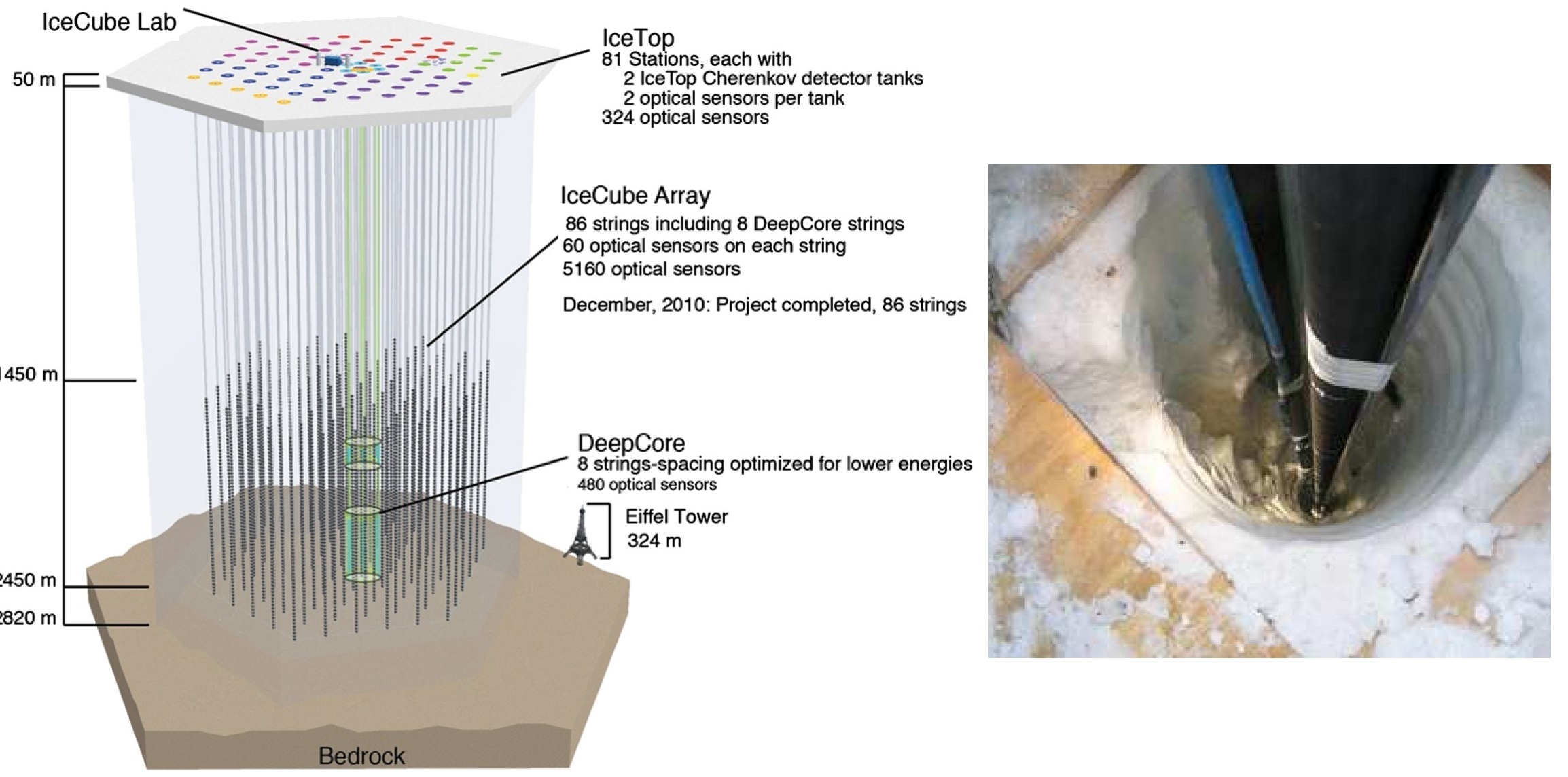}
\end{center}
\caption{\small Schema del rivelatore Icecube al Polo Sud. Oltre una matrice di rivelatori sotto il ghiaccio, sono visibili anche dei rivelatori posti sopra la superficie. Nella foto a destra: uno dei buchi perforati nel ghiaccio con l'inserimento della stringa di PMT.}
\label{fig:icecube}
\end{figure}

L'esperimento IceCube (Fig. \ref{fig:icecube}), condotto da una vasta collaborazione internazionale, è stato installato sotto i ghiacci del Polo Sud tra il 2005 ed il 2010, a una profondità che va da 1450 a 2450 metri. 
Utilizza più di 5000 PMT, distribuiti su 86 stringhe lunghe circa 1000 m. Le stringhe sono state posizionate dopo aver creato buche nel ghiaccio con getti di acqua calda. 
La posizione delle stringhe è quindi fissata una volta per tutte e nessuna operazione di recupero/riparazione è possibile. Non sono necessarie (a differenza di esperimenti sottomarini) procedure di calibrazione per il posizionamento dei PMT. Inoltre il ghiaccio ha una trasparenza maggiore di quella dell'acqua, ha una contaminazione di elementi radioattivi trascurabile rispetto il mare e non ha bioluminescenza.
Tuttavia, minuscole bolle presenti nel ghiaccio aumentano la probabilità che la luce Cherenkov sia deflessa, peggiorando rispetto l'acqua la precisione nella misura della direzione delle tracce dei muoni.

IceCube ha recentemente osservato per la prima volta neutrini di origine astrofisica.
Come abbiamo detto, i neutrini atmosferici (originati cioè dall'interazione dei raggi cosmici con i nuclei dell'atmosfera) rappresentano un fondo irriducibile.
Tuttavia, ci si aspetta che il numero di neutrini atmosferici decresca con l'energia come $\sim E^{-3.7}$, mentre i neutrini di origine cosmica (come abbiamo visto) decrescono come $\sim E^{-2}$.
Quindi, oltre una certa energia di soglia, il numero di neutrini cosmici dovrà essere in eccesso rispetto a quelli atmosferici. 
Si tratta quindi di individuare e utilizzare qualche grandezza osservabile che permetta una stima dell'energia del neutrino incidente (\textit{proxy} dell'energia), e confrontare quanti eventi atmosferici sono attesi, e quanti sono misurati. Un eccesso negli eventi misurati è una indicazione di un flusso di neutrini extraterrestri.

\begin{figure}[tbh]
\begin{center}
\includegraphics[width=12.5cm]{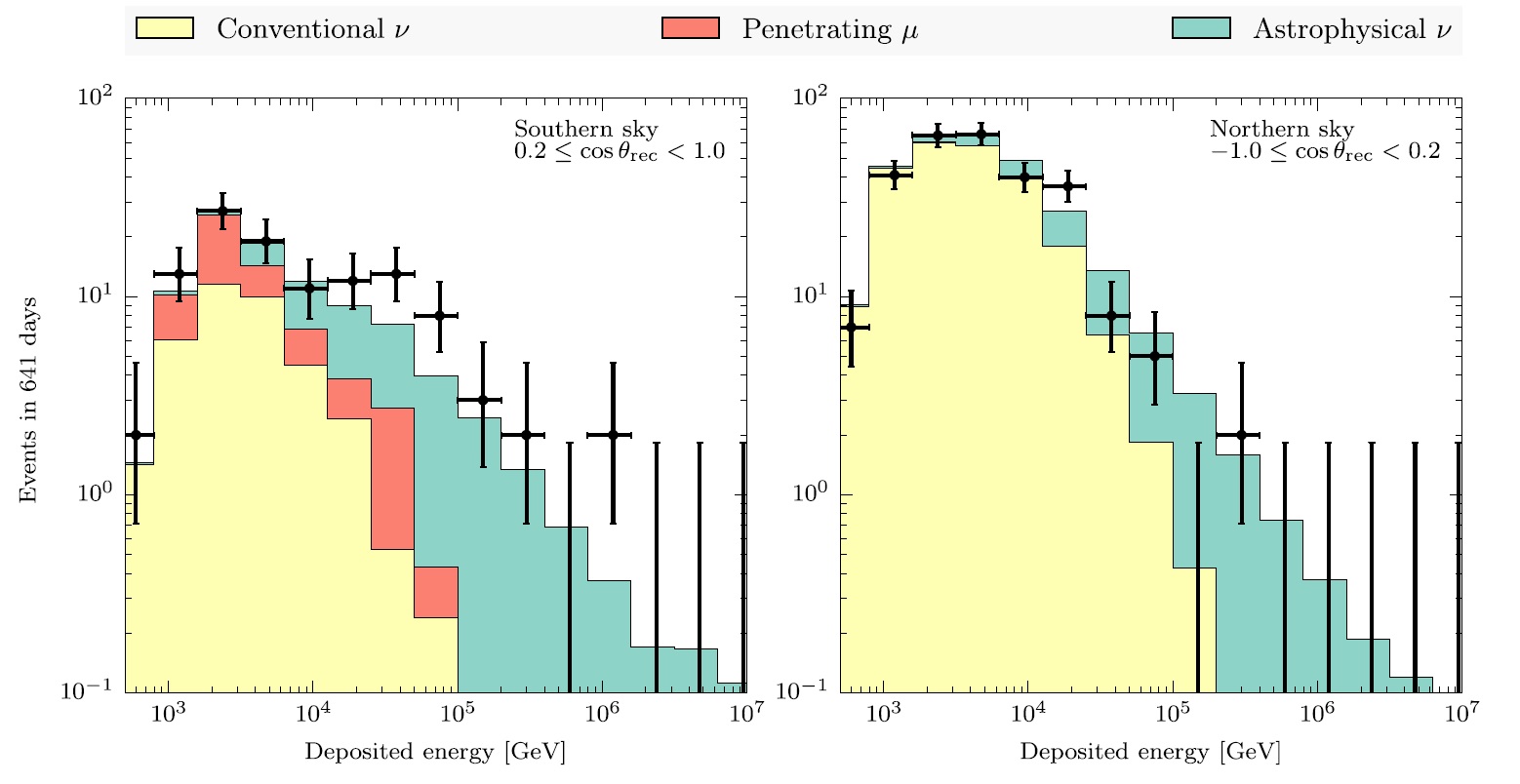}
\end{center}
\caption{\small 
Spettro dell'energia depositata dagli eventi selezionati nel volume fiduciale di IceCube. A sinistra (destra) gli eventi dall'emisfero Sud (Nord) del cielo, ricostruiti come eventi diretti verso il basso (l'alto).
I punti in nero, con barra d'errore statistico, sono i dati. Il numero di eventi aspettati dai neutrini atmosferici sono rappresentati con l'istogramma in giallo; i muoni atmosferici con l'istogramma arancio. Sopra $\sim 10^4$ GeV è necessario aggiungere una componente di neutrini cosmici per spiegare i dati, rappresentata dall'istogramma in verde. La dipendenza energetica trovata per questi ultimi è del tipo $\sim E^{-2.5}$ \cite{ic-hese}. }
\label{fig:hese}
\end{figure}

IceCube ha evidenziato un eccesso di eventi sopra il fondo dei neutrini atmosferici in due campioni indipendenti di eventi.
Il primo campione (denominato HESE, ossia \textit{High Energy Starting Events}) è rappresentato dai neutrini che interagiscono in una regione fiduciale del rivelatore. Gli strati esterni di IceCube sono usati come \textit{veto} per i muoni atmosferici. 
Un evento che ``miracolosamente appare'' nel rivelatore senza essere visibilmente entrato rappresenta infatti l'arrivo di una particella neutra, che interagisce poi all'interno del rivelatore trasformandosi in un leptone carico.
In questa tipologia di eventi, nei dati raccolti nel periodo 2010-2012 e sinora pubblicati (Fig. \ref{fig:hese}), si evince un eccesso di eventi quando l'energia depositata è $>$ 30 TeV.
La gran parte di questi eventi sono del tipo $\nu_e \rightarrow e \rightarrow$ cascata elettromagnetica, ed hanno una precisione nella ricostruzione angolare di soli 10$^\circ$-15$^\circ$. 
Probabilmente anche per questo motivo, non si è evidenziato nel campione sinora raccolto nessun eccesso rispetto una ipotesi di segnale isotropo.

\begin{figure}[tbh]
\begin{center}
\includegraphics[width=9.5cm]{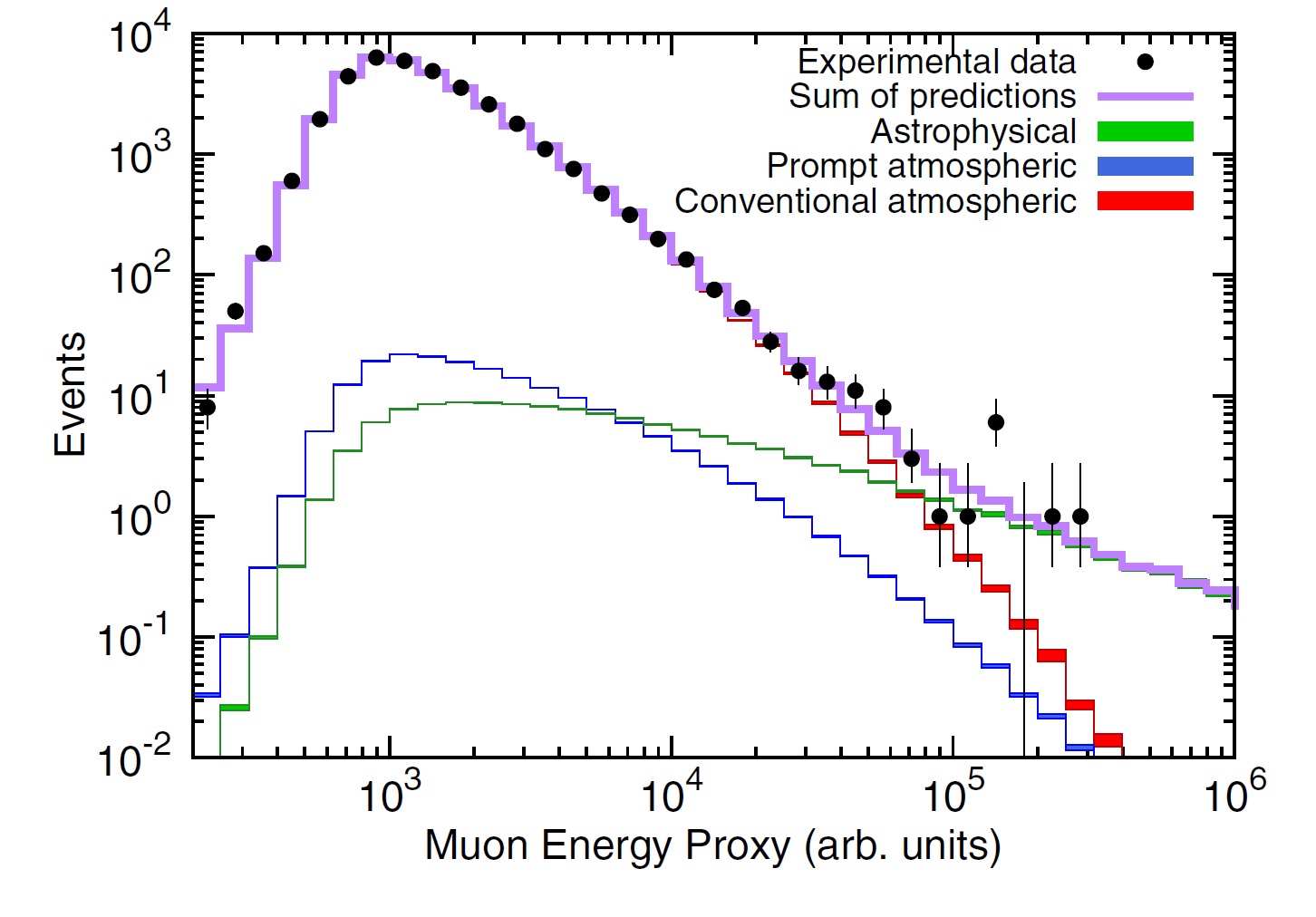}
\end{center}
\caption{\small Spettro in funzione dello stimatore dell'energia dei $\mu$ verso l'alto (quindi provenienti dal cielo Nord) misurati in IceCube.
I punti in nero, con barra d'errore statistico, sono i dati. In rosso il contributo atteso dai neutrini atmosferici. In blu quello dei neutrini atmosferici prodotti dal decadimento di adroni pesanti. In verde il contributo calcolato per i neutrini di origine astrofisica, avendo ipotizzato una dipendenza dall'energia del tipo $E^{-2}$. }
\label{fig:ICmu}
\end{figure}

Una seconda tipologia di eventi in cui IceCube ha evidenziato un eccesso di eventi è quella che utilizza $\nu_\mu \rightarrow \mu \rightarrow$ traccia passante \cite{ic-mu}. Questi eventi sono solamente diretti verso l'alto e quindi (vista la posizione di IceCube al Polo Sud) provenienti dall'emisfero Nord del cielo. Anche in questo caso si evidenzia un significativo eccesso di eventi di alta energia rispetto l'atteso dei neutrini atmosferici, Fig. \ref{fig:ICmu}. 
La precisione nella determinazione della direzione originaria del neutrino è, in questo caso, dell'ordine di 1$^\circ$. Tuttavia, anche in questo campione di eventi non si evidenzia nessuna concentrazione da una particolare direzione del cielo.

Si noti che tra i due campioni di eventi astrofisici vi è una discrepanza nella dipendenza dall'energia, che viene ottenuta dal migliore adattamento dei dati con una curva teorica del tipo $E^{-\Gamma}$ (il parametro $\Gamma$ viene chiamato \textit{indice spettrale}).
Nel primo campione di Fig. \ref{fig:hese}, gran parte degli eventi sono del tipo cascata elettromagnetica, e il neutrino originario può provenire sia dal cielo Nord che da quello Sud. La dipendenza con l'energia ricostruita ha indice spettrale $\Gamma \simeq {2.5}$. 
Nel caso dei muoni verso l'alto di Fig. \ref{fig:ICmu}, gli eventi  sono provenienti dal cielo Nord, e il migliore adattamento con l'energia ricostruita ha indice  $\Gamma \simeq {-2.0}$.
Per questa discrepanza, sono possibili diverse interpretazioni: 
\begin{itemize}
\item essa è semplicemente dovuta a una fluttuazione statistica, che scomparirà all'aumentare del numero di eventi;
\item è un effetto del fatto che nel primo campione sono dominanti i $\nu_e$ (eventi $cascata$) e nel secondo i $\nu_\mu$ (eventi $traccia$);
\item è un effetto dovuto al fatto che nel primo campione vi sono eventi sia dal cielo Nord che dal cielo Sud, mentre nel secondo solo eventi dal cielo Nord.
\end{itemize}
L'ultima ipotesi è suffragata dal fatto che il piano della Via Lattea è situato nel cielo Sud. 
L'emisfero Nord contiene solo una piccola parte del piano galattico.
Quindi, mentre i neutrini provenienti dell'emisfero Nord potrebbero essere di origine principalmente extra-galattica (con indice spettrale $\Gamma\simeq 2.0$), i neutrini provenienti dal cielo Sud potrebbero avere una componente mista, galattica ed extra-galattica.
La sovrapposizione di due componenti potrebbe portare ad un eccesso di eventi di origine astrofisica con diverso indice spettrale rispetto ai neutrini extra-galattici.

La discriminazione tra le tre menzionate ipotesi potrà risolversi solo con l'aumento del numero di eventi raccolti da IceCube e, soprattutto, con gli eventi misurati dal telescopio in costruzione KM3NeT.
Questo infatti avrà intrinsecamente una sensibilità almeno paragonabile a quella di IceCube, una migliore visibilità dell'emisfero Sud celeste (col piano della Galassia) e una migliore risoluzione angolare per la misura degli eventi di tipo traccia.

I prossimi dieci anni saranno quindi decisivi per l'astrofisica dei neutrini di alta energia, per cercare di capire quali sono i meccanismi capaci nell'universo di accelerare particelle a energie straordinariamente elevate. Questo è uno dei misteri per l'astrofisica attuale.

\section{Ulteriori ``miracoli'' del neutrino. Le oscillazioni e cosa significano\label{sec:osci}}

Torniamo un attimo sulla Terra prima di terminare. 
Oltre a contribuire in maniera decisiva allo studio dell'astrofisica stellare, a quello del collasso gravitazionale stellare e (nel prossimo futuro) a quello degli oggetti stellari collassati e dell'astrofisica delle alte energie, lo studio dei neutrini di origine naturale ha contribuito, tra il 1998 e il 2002, alla scoperta delle \textit{oscillazioni dei neutrini}, che ha avuto un impatto significativo sul modello standard del microcosmo.
La scoperta delle oscillazioni è avvenuta quasi contemporaneamente con lo studio dei neutrini atmosferici e quelli solari, ed è stata successivamente confermata con esperimenti che utilizzano neutrini originati da acceleratori di particelle (dal CERN in Europa, da FermiLab negli USA e dai laboratori J-Park e KEK in Giappone) e da reattori nucleari (che, come abbiamo visto, sono potenti sorgenti di $\overline \nu_e$).

Nel modello standard del microcosmo i tre neutrini \( \nu _{e},\nu _{\mu },\nu _{\tau } \) hanno massa nulla, e appartengono (per usare un gergo diverso) a tre diverse \textit{caste}: una volta che sei neutrino elettronico resti neutrino elettronico per l'eternit\`{a}. 
Tecnicamente, questo viene chiamato \textit{conservazione dei numero leptonico di famiglia (o di sapore)}, legge che deve essere introdotta \textit{ad hoc} nella teoria. 
Inoltre, il fatto che tutti e tre i sapori di neutrini abbiamo massa nulla è sorprendente (è una situazione che in fisica quantistica viene chiamata \textit{degenerazione}).
Negli anni '60 (partendo da una idea di Bruno Pontecorvo) si iniziò ad ipotizzare la possibilit\`{a} di oscillazioni di un neutrino di un certo sapore in un neutrino di sapore differente.
Questo \`{e} un fenomeno possibile nella teoria solo se i neutrini hanno massa non nulla e differente tra i sapori.

La spiegazione (e comprensione) delle oscillazioni dei neutrini necessita delle conoscenze di fisica quantistica. 
Possiamo per\`{o} provare a semplificare, dicendo innanzitutto che in realt\`{a} non \`{e} corretto parlare di massa dei neutrini \( \nu _{e},\nu _{\mu },\nu _{\tau } \).
I neutrini che chiamiamo \( \nu _{e},\nu _{\mu },\nu _{\tau } \) sono quelli che appaiono (scompaiono) quando scompare (appare) un $e$, $\mu$ o un $\tau$ (come abbiamo studiato nelle sezioni precedenti). 
Sono quindi chiamati \textit{{}``autostati d'interazione}''.
Una volta che un neutrino viene prodotto, \`{e} libero di propagarsi.
Se la massa \`{e} nulla, i neutrini \textbf{debbono} viaggiare alla velocit\`{a} della luce, $c$.
Ma se hanno massa, ancorché piccola, debbono viaggiare a una velocit\`{a} minore di $c$\footnote{Nel 2011 vi fu una eccitazione mai vista nel mondo della fisica. Un esperimento che rivelava al Gran Sasso fasci di neutrini sparati dal CERN, a 730 km di distanza, riportò con gran rumore mediatico che i neutrini ``viaggiavano'' più veloci della luce. Dopo qualche mese di subbuglio, qualche membro dell'esperimento ha realizzato che una cattiva connessione di una fibra ottica provocava un ritardo artificiale del segnale di sincronizzazione temporale CERN-Gran Sasso che si propaga alla velocità della luce. Non erano i neutrini ad anticipare, ma la luce a ritardare per un errore tecnico.} e che dipende dalla massa (maggiore \`{e} la massa, minore la velocit\`{a} a parit\`{a} di energia). 
\begin{figure}[tbh]
\begin{center}
\includegraphics[width=7.5cm]{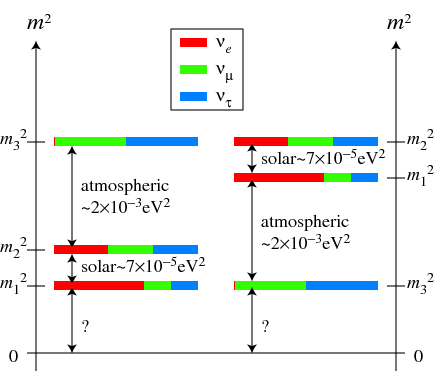}
\end{center}
\caption{\small Rappresentazione schematica del miscelamento tra \textit{autostati di massa} dei neutrini (indicati con $\nu _{1},\nu _{2},\nu _{3}$) e autostati di sapore (indicati coi tre diversi colori per $\nu _{e},\nu _{\mu },\nu _{\tau }$). Quello che si è misurato con gli esperimenti di oscillazione dei neutrini è la differenza tra le masse. 
Gli esperimenti con i neutrini atmosferici (Super-Kamiokande, MACRO e altri) hanno misurato la differenza tra i quadrati delle masse 
$m^2_{3}-m^2_{2}=2.3\times 10^{-3}$ eV$^2$. 
Gli esperimenti per la misura dei neutrini solari (Gallex, SK, SNO, Borexino ed altri)  hanno misurato la differenza tra i quadrati delle masse 
$m^2_{2}- m^2_{1}=7.6\times 10^{-5}$ eV$^2$. Non si conosce il valore assoluto della massa più bassa. La parte sinistra della figura rappresenta la \textit{gerarchia normale}, quella di destra la  \textit{gerarchia inversa} (vedi testo).}
\label{fig:numass}
\end{figure}

Se esistono tre famiglie, possono esistere tre possibili diversi valori di massa. I neutrini con masse diverse vengono chiamati \textit{autostati di massa} e indicati con \( \nu _{1},\nu _{2},\nu _{3} \).
Nulla vieta che si possano avere (ad esempio) le seguenti relazioni: $\nu_e=\nu_1$, $\nu_\mu=\nu_2$ e $\nu_\tau=\nu_3$. 
\textbf{Oppure}, di avere combinazioni lineari (cioè, miscelamenti) tra autostati d'interazione e autostati si massa. 
Ad esempio, potrebbe essere che 
$\nu_1= 60\% \nu_e + 21 \% \nu_\mu + 19\% \nu_\tau$ e analogamente per $\nu_2$ e $\nu_3$ (si veda la Fig. \ref{fig:numass}).
Ovviamente, un sistema lineare di tre equazioni e tre incognite si può invertire, e quindi possiamo ad esempio dire che il neutrino elettronico (rappresentato dal colore rosso in figura) è in maggior parte $\nu_1$, in minoranza $\nu_2$ e proprio una traccia di $\nu_3$, ossia (approssivativamente)
$\nu_e= 62\% \nu_1 + 36 \% \nu_2 + 2\% \nu_3$.

Cosa significa questa situazione? 
Il neutrino nasce (ad esempio) come neutrino elettronico, ma quando si propaga si propaga come una miscela di tre diversi stati, in cui domina (al 62\%) l'autostato di massa $m_1$. 
Ma le tre diverse componenti si propagano con diverse velocità (perché hanno diverse masse), e quindi in un punto lontano e ad un certo istante di tempo la particella si presenta \textit{diversa} da come era stata prodotta nel punto di partenza. Se potessimo rivelarla, ovvero ottenere una interazione che produce un leptone carico, potremmo NON rivelare un elettrone ma, con una certa probabilità, un muone o un tau, se l'energia del neutrino è sufficiente per formarne la massa a riposo.

Serviamoci di una analogia con l'ottica.
In ottica distinguiamo fra ``colori base'' (rosso, blu e giallo) e
``colori composti''. Ad esempio, il violetto \`{e} un miscuglio di rosso e di blu. 
Immaginiamo che una certa sorgente \textit{generi} un'onda {}``violetta''. Il violetto (corrispondente nell'analogia coi neutrini ad un autostato d'interazione, che potrebbe corrispondere ai $\nu_e$ prodotti dalle reazioni all'interno del Sole) \`{e} in realt\`{a} un colore composto, formato dal mescolamento dei colori base (corrispondenti agli autostati di massa) rosso e blu. 
L'onda emessa (=$\nu_e$ dal Sole) \`{e} quindi composta da un'onda rossa e da una blu con valori iniziali tali da dare, nel loro miscuglio, la giusta tonalit\`{a} di violetto. 
Per la \textit{propagazione}, conside\-ria\-mo invece i colori base rosso e blu. Se le onde rossa e blu si propagano con la stessa velocità, la loro sovrapposizione d\`{a} ovunque lo stesso colore violetto. 
Se invece si propagano con velocit\`{a} diversa
\footnote{Si noti che la luce si propaga con velocità $v$ differente per ogni colore in un mezzo (si pensi al prisma di Newton) perché l'indice di rifrazione $n(\lambda)$ dipende dalla lunghezza d'onda, e $v=c/n(\lambda)$. Per i neutrini che si propagano nel vuoto si ha una relazione analoga, $v=c/n(m)$, dove $n(m)$ è una funzione che dipende dalla massa $m$ del neutrino.}, la loro proporzione \`{e} diversa da punto a punto, e parimenti lo \`{e} il colore risultante \textit{visto} dall'osservatore, il cui occhio \`{e} globalmente sensibile non ai colori base isolati, ma al loro miscuglio o sovrapposizione.
Il fatto che il colore di partenza sia in realt\`{a} composto da due diversi colori base (autostati di massa) e che questi si propaghino diversamente, d\`{a} luogo all'osservazione di un colore composto
(autostato d'interazione), diverso da quello di partenza e variabile da punto a punto. 
La parola {}``oscillazione'' non si riferisce, in effetti, al fatto che le particelle sono rappresentate da onde, ma piuttosto al fatto che il colore osservato (autostato d'interazione) cambia allontanandosi dalla sorgente, con legge oscillatoria. In certi punti, l'onda potr\`{a} apparire ad un osservatore addirittura come puramente rossa o blu.

Le conseguenze delle oscillazioni dei neutrini sulla nostra conoscenza del microcosmo e del macrocosmo non sono ancora state tutte chiarite. 
Per esempio, c'è chi sospetta che l'asimmetria tra materia e antimateria osservata nell'universo (uno dei grandi enigmi della cosmologia) possa essere una conseguenza delle proprietà dei neutrini. 
Sono attualmente in fase di progettazione esperimenti per proseguire nello studio della natura dei neutrini, facendo uso anche di fasci di neutrini artificialmente prodotti tramite acceleratori.

\section{Conclusioni\label{sec:conc}}

Il neutrino è probabilmente la particella che offre le maggiori e più evidenti connessioni tra fisica delle particelle, astrofisica e cosmologia.
Potrete trovare ulteriori informazioni e spunti per approfondimenti in una serie di articoli in lingua italiana pubblicati in una rassegna completamente dedicata ai neutrini \cite{Ithaca}. 
Nei prossimi 10-20 anni quattro saranno i campi di indagine principali riguardanti il neutrino. 

Il primo riguarda l'astrofisica delle alte energie. 
IceCube ha infatti evidenziato l'esistenza di neutrini di origine cosmica, ma non è sinora stato in grado di capire da quali oggetti questi neutrini sono stati originati (sorgenti galattiche, extragalattiche).
Il nuovo telescopio di neutrini KM3NeT in avanzata fase di realizzazione nel mar Mediterraneo, a largo delle coste siciliane, ha la capacità di aiutare a risolvere in maniera chiara il problema.

Il secondo riguarda quello che si chiama \textit{il problema della gerarchia di massa}. Nella parte sinistra di Fig. \ref{fig:numass} è rappresentata schematicamente la situazione in cui il neutrino $\nu_1$ ha massa più bassa e quello $\nu_3$ ha massa più elevata. Questo ci sembra l'ordine naturale (1 viene prima di 2 ed entrambi prima di 3), e la situazione viene chiamata \textit{gerarchia normale}.
Ma avere chiamato i neutrini 1, 2 e 3 è solo una nostra convenzione. Quello che importa sono solo le \textbf{misure}. E le misure hanno permesso di stabilire le differenze tra le masse dei diversi autostati, e che la massa $m_1$ è più piccola di $m_2$. Quindi, potrebbe essere che l'autostato $m_3$ sia nel gradino più basso, $m_1$ nella posizione intermedia e $m_2$ in alto! Questa configurazione (parte destra della figura) viene chiamata \textit{gerarchia inversa}.
Diversi progetti sono in competizione per chiarire questa situazione dal punto di vista sperimentale nei prossimi anni, utilizzando neutrini atmosferici (KM3NeT-ORCA in Francia, Pingu al Polo Sud) oppure neutrini da reattori nucleari (JUNO in Cina). A meno che, non vi sia una esplosione di supernova nella nostra Galassia che contribuisca a definire la situazione.

È importante questa distinzione tra gerarchia normale e inversa? È molto importante per capire la natura del neutrino.
Questo rappresenta il terzo campo di indagine degli esperimenti in corso.
Il modello standard del microcosmo prevede che per ogni sapore (ad esempio, $\nu_e$) vi siano due stati distinti legati allo spin della particella.
Per far una analogia meccanica (non del tutto corretta), ogni particella può essere assimilata ad una piccola trottola che può ruotare in senso orario oppure antiorario. 
Sperimentalmente, tutti i leptoni carichi e i quark rispettano queste caratteristiche, che sono formalizzate nella teoria da una serie di equazioni quantistiche dette di Dirac (equazioni scritte originariamente per descrivere l'elettrone) come inglobato nel Modello Standard del microcosmo.
Per il neutrino la situazione sembra diversa: infatti, sperimentalmente, è noto solo il $\nu_e$ che (diciamo) ruota in senso antiorario, mentre non è mai stato osservato il $\nu_e$ che ruota in senso orario. Viceversa, è stato osservato solo il $\overline \nu_e$ che ruota in senso orario, e mai in senso antiorario.
Per quale motivo mancano questi stati descritti dalle equazioni di Dirac?

Un tentativo di spiegazione venne dato molto tempo fa da un fisico italiano, Ettore Majorana\footnote{La fama di Majorana tra i fisici è dovuta a questa ipotesi. Tra i non fisici Majorana è più noto per essere salito a Palermo su un traghetto e non essere mai disceso a Napoli quando il traghetto ha attraccato.}. 
Majorana ipotizzò che il neutrino e antineutrino siano la stessa particella; noi erroneamente chiamiamo particella e antiparticella i due stati di spin di questo unico oggetto (la rotazione oraria e quella antioraria). 
Se il neutrino è una particella di Majorana, il modello standard del microcosmo dovrà essere rivisto in maniera significativa. Esso infatti ingloba solo particelle di tipo Dirac. Modificare il modello standard non significa solo cambiare delle formule di matematica: significa introdurre nuovi concetti, nuovi meccanismi con cui le particelle possono interagire tra di loro.

Chi può decidere se il neutrino è di Dirac o di Majorana?
Ovviamente, gli esperimenti. Molti sono in corso nei laboratori sotterranei, al Gran Sasso e in altre parti del mondo.
Tuttavia, una cosa la sappiamo: se la gerarchia del neutrino è normale (come rappresentato a sinistra di Fig. \ref{fig:numass}), la risposta sperimentale a questa domanda potrebbe essere molto lontana nel tempo.
Nel caso in cui, invece, la gerarchia del neutrino è inversa, la risposta potrebbe aversi anche nel prossimo decennio.

L'ultimo campo di indagine, infine, è quello da rappresentato dallo studio delle proprietà dei neutrini, confrontate con quelle degli antineutrini. Se neutrini e antineutrini avessero piccole differenze nelle loro proprietà d'interazione, queste potrebbero spiegare l'asimmetria materia/antimateria nell'universo visibile, citata in conclusione del paragrafo precedente.

\vskip 0.2truecm
La fisica delle particelle e l'astrofisica hanno ancora molte domande; a molte possiamo ottenere la risposta solo tramite lo studio dei neutrini.
Come il poeta, cerchiamo nell'oscurità di udire i racconti di una novella Francesca da Rimini.

\vskip 0.4 truecm
\noindent \textbf{Ringraziamenti}

\noindent L'autore ringrazia la presidenza del Centro Fermi e il dr. R. Nania dell'INFN-Bologna per l'invito alla conferenza che ha stimolato il presente articolo.
Inoltre ringrazia vivamente il prof. Francesco Vissani del GSSI e la prof.ssa Mariagrazia Fabbri del Liceo Galvani di Bologna per utilissimi commenti alle bozze.  


\end{document}